\documentclass[12pt,reqno]{article}
\usepackage[lmargin=2cm, rmargin=2cm]{geometry}
\usepackage[dvips]{graphicx}
\usepackage{amssymb}
\usepackage{appendix}
\usepackage{datetime,color}

\usepackage[utf8]{inputenc}
\usepackage[english]{babel}
\usepackage{amsfonts, color}
\usepackage{amsmath, amssymb, graphics}
\usepackage{mathrsfs}
\newcommand{\mathsym}[1]{{}}

\usepackage{cancel, slashed}

\newcommand{\g}{\gamma}

\newcommand{\mc}{\mathcal}
\newcommand{\bb}{\mathbb}
\newcommand{\p}{\partial}

\newcommand{\rar}{\rightarrow}

\newcommand{\nn}{\nonumber}
\newcommand{\e}{\epsilon}

\newcommand{\im}{\mathrm{Im}}
\newcommand{\re}{\mathrm{Re}}

\def\bfone{\relax{\rm 1\kern-.35em 1}}

\makeatletter \@addtoreset{equation}{section} \makeatother

\usepackage{fullpage}
\usepackage{hyperref}

 \allowdisplaybreaks[3]

\title{\vspace{-0.5cm} On the non-BPS extension of first order flow \\ in $\mc N=2$ $U(1)$-gauged Supergravity \vspace{-1.1cm}}

\begin{document}

\begin{titlepage}
 \thispagestyle{empty}
 \begin{flushright}
     \hfill{ITP-UU-12/40}\\
\hfill{SPIN-12/37}
 \end{flushright}

 \vspace{50pt}

 \begin{center}
     { \Large{\bf      {On the non-BPS first order flow \\ [5mm] in $\mc N=2$ $U(1)$-gauged Supergravity }}}

     \vspace{60pt}

     {\large { Alessandra Gnecchi\,\, {\small\&}\,  Chiara Toldo}}
\\[15mm]
{\slshape
Institute for Theoretical Physics \emph{and} Spinoza Institute, \\[3mm]
Utrecht University, 3508 TD Utrecht, The Netherlands \\[4mm]

{\upshape\ttfamily  A.Gnecchi@uu.nl,  C.Toldo@uu.nl}\\[3mm]}

\vspace{8mm}

\vspace{20pt}

     \vspace{20pt}

    \vspace{20pt}

     \vspace{20pt}

     {\bf Abstract}
 \end{center}

 \vspace{10pt}
\noindent 
We consider theories of $\mathcal{N}=2$ supergravity with Fayet-Iliopoulos gauging and describe a procedure to obtain non-BPS extremal black hole solutions in asymptotically AdS$_4$ space, in a fully symplectic covariant framework. 

By considering both electric as well as magnetic gauging, we are able to find new extremal purely magnetic and dyonic solutions.  We consistently impose the Dirac quantization condition as a constraint on the black hole and gravitinos charges. This additional requirement allows to parametrize the black hole entropy in terms of an integer and of the entropy of the corresponding black hole in the ungauged model. 

We also find the nonextremal generalization of the dyonic solution and we compute the product of the areas. For all the configurations with asymptotic supersymmetry we furthermore compute the mass. 
 \vfill

\end{titlepage}

\baselineskip 6 mm

\tableofcontents

\section{Introduction and Outlook}

There has been lately some effort in characterizing AdS black hole solutions in gauged supergravity. These solutions, BPS and non-BPS, are in general less known than their cousins in ungauged supergravity.

Although black holes in gauged supergravities have been known for a long time \cite{Sabra:1999ux}\nocite{Chamseddine:2000bk}\nocite{Caldarelli:1998hg}\nocite{Duff:1999gh} \nocite{Cucu:2003yk}\nocite{Bellucci:2008cb}-\cite{Kimura:2010xe}, all supersymmetric solutions in these theories were thought to have vanishing horizon. Given the lack of a regular extremal configurations, properties such as the entropy-area law have not been extensively studied, and no further investigations on the zero temperature configurations have been carried out.
Only recently, in fact, it has been shown that it is possible to have genuinely finite horizon black holes if one reduces the supersymmetry preserved by the solution to be 1/4 of the original vacuum \cite{Cacciatori:2009iz,Dall'Agata:2010gj,kiril_stefan}. This now opens  the possibility of getting important insights on the physics of black holes in gauged supergravity, such as the microstate structure of the entropy of extremal black holes.

In the BPS sector,  one can construct dyonic black holes with spherical horizon (finite nonzero area of the event horizon). They can be deformed to nonextremal ones \cite{ours_nonextr,klemm}, in order to have thermal states, which are useful for applications of AdS/CFT to condensed matter systems. 
Moreover, these finite temperature black holes provide another playground where to test the conjecture of \cite{Larsen:1997ge,Cvetic:2010mn,Galli:2011fq,Castro:2012av} concerning the product of the inner and outer areas of the horizons. Indeed, for all the non-extremal cases considered so far, such product does not depend on the mass of the configuration, but only on the quantized charges.

Many Supergravity models can be regarded as low energy limit of String or M-theory, and, in these cases, black holes solutions correspond to configurations of fluxes and branes. In particular, gauged supergravities are obtained upon suitable compactifications with fluxes, that source a potential in the low energy theory. 
Given the importance of fluxes in addressing the issue of moduli stabilization, it is crucial to study the attractor mechanism in gauged supergravity, that might destabilize the string theory vacuum. 
The supersymmetric solutions are subject indeed to a  \lq{}\lq{}double attractor\rq{}\rq{} condition,  meaning that supersymmetry fixes the value of the scalars both at asymptotic infinity and at the horizon, and can be in conflict with the minimization of the potential generated by the flux compactification.

In ungauged supergravity, in addition to supersymmetric configurations, there exists extremal solutions which break all supersymmetry, and nonetheless obey a first order flow. 

The main aim of this paper is to present a way to get extremal but non-BPS solutions starting from BPS ones. The procedure is based on the one already studied in the ungauged case \cite{Ceresole:2007wx}, and consist in a symplectic rotation of the charges of the BPS configuration. This procedure enlarges the zoo of solutions at our disposal providing new examples of genuine extremal black holes. They are required to satisfy a Dirac quantization condition due to the fact that, in the presence of gauging, the gravitinos are charged:
\begin{equation}\label{dirac_ultima}
g_{\Lambda}p^{\Lambda}-g^{\Lambda}q_{\Lambda}=n\qquad n \in \mathbb{Z}\,.
\end{equation}
 In the supersymmetric case, this condition is automatically satisfied and supersymmetry picks out just the values $\pm 1$.
As we will see, for the non-BPS configurations we have to impose by hand the Dirac quantization condition. This provides us with a tower of extremal black holes for any integer $n$.

We apply this solution generating technique to the solutions found in \cite{Dall'Agata:2010gj} and \cite{kiril_stefan}, namely for dyonic and purely magnetic black holes in presence of mixed and electric gaugings, respectively. 
They are actually equivalent configurations, since the frames in which they are constructed can be transformed one into the other by a symplectic transformation, followed by a holomorphic coordinate redefinition.
We then deform the dyonic solution to a nonextremal one, generating a new thermal black hole that is regular in the extremal limit. For this new solution, the product of the areas is also verified to be independent of the mass.
 
All these extremal and nonextremal configurations provide a new piece of information about the spectra of the solutions in gauged supergravity, and at the same time raise a lot of challenging questions, related, e.g., to the thermodynamical aspects of stability for non-BPS solutions. Some subtleties appear when one tries to compute the mass of these solutions. In fact, the procedure described in \cite{ours_BPSbound} and \cite{kiril} requires that the configuration preserves some supersymmetry at least asymptotically, and for the non-BPS solutions presented here this is not the case. Finally, given also recent developments \cite{Hristov:2012nu}, it would be important to  investigate further the relation between black holes in ungauged and gauged supergravity, and to study which properties of the former generalizes to the latter case.

\hspace{1mm}

\textbf{Note added}: During the write-up of our work, the paper
arXiv:1211.1618 by D. Klemm and O. Vaughan appeared \cite{klemm4}. Their work present some overlap with
our analysis for what concerns the technique for generating extremal non supersymmetric black holes.

\section{\label{trick}non-BPS rotation trick}

It has been shown long ago \cite{Ceresole:2007wx} that, for ungauged supergravity theories, it is possible to obtain extremal black holes solutions by a suitable symplectic rotation of the charges of a BPS configuration, and thus derive a fake superpotential that drives the first order non supersymmetric flow. 
More explicitly, such rotation acts linearly on the charges as a constant matrix $S\in Sp(2n_V+2,\mathbb R)\ $\footnote{We denote by $n_V$ the number of abelian vector multiplets of the $N=2$ theory under consideration. Together with the graviphoton, the theory has a total of $n_V+1$ abelian gauge fields, and the duality group $G$ is a subset of $Sp(2n_V+2, \mathbb R)$ \cite{Gaillard:1981rj}.}, that does not act as a duality transformation, in particular it only affects the charges and not the scalar fields. It is only a tool to achieve a different squaring of the action and thus get to a set of first order non-BPS equations, in the same way as for the ungauged Supergravity case. There, the same rotation $S$ was first introduced by Ceresole and Dall'Agata (see Sec. 3 of \cite{Ceresole:2007wx}).
In particular, some non-BPS black holes can be derived by simply flipping some signs of the charges of the BPS solution. 

We are going to show that the same conceptual idea also works for $N=2$ Supergravity with $U(1)$-gauging. This turns out to be particularily straightforward, since the only additional contribution to the Lagrangian is the gauging potential $V_g$. 
To make the derivation and the references clearer, we first review the mechanism that leads to the BPS squaring of the action, starting from the one dimensional effective action derived in \cite{Dall'Agata:2010gj}.

\subsection{\label{BPS-squaring-L}The original BPS squaring of the action}
We consider models of $N=2$ Supergravity, in presence of electric and magnetic gauging $\mc G=( g^\Lambda,g_\Lambda)$, along the lines of \cite{Dall'Agata:2010gj}. In the same paper, the first order equations of motion were derived for extremal supersymmetric configurations, by \lq{}\lq{}completing the square" of the terms in the effective one dimensional action. Let us review how the BPS squaring works for this theory. We refer to \cite{Dall'Agata:2010gj} for further notations and conventions.


The bosonic Lagrangian of the $U(1)$-gauged theory is 
\begin{equation}
	{\mathscr L} = \frac{R}{2} - g_{i\bar\jmath}\, \partial_\mu z^i \partial^\mu \bar z^{\bar \jmath} + \frac14\,\hbox{Im}{\cal N}_ {\Lambda\Sigma}\, F^\Lambda_{\mu \nu}\, F^{\Sigma\,\mu\nu} +\frac14\,\hbox{Re}{\cal N}_{\Lambda\Sigma}\, F^{\Lambda}_{\mu\nu}\,\frac{\epsilon^{\mu\nu\rho \sigma}}{2\sqrt{-g}} F^\Sigma_{\rho \sigma} - V_g \, ,
	\label{StartingAction} 
\end{equation}
where the indices $i,\bar \jmath$ for the scalar fields run from 1 to $n_V$ and the symplectic indices are $\Lambda, \Sigma =1,..., n_V+1$. The only difference with respect to the Lagrangian of the ungauged theory is the scalar potential $V_g$ \cite{Andrianopoli:1996cm}. This is generated by the FI  terms $\cal G$ and can be written in a symplectic covariant form as
\begin{equation}
	V_g = g^{i\bar\jmath}\,  D_i {\cal L} \overline{D}_{\bar \jmath} \overline{\cal L} - 3 |{\cal L}|^2 \qquad \left(\hbox{where}\quad D_i {\cal L} \equiv \partial_i {\cal L} + 1/2\, \partial_i K \, {\cal L}\right)
	\label{Vg}
\end{equation}
in terms of a superpotential 
\begin{equation}
	{\cal L} =  \langle {\cal G}, {\cal V}\rangle = \mc G^T \Omega \mc V
=e^{K/2} \left(X^\Lambda g_{\Lambda}- F_{\Lambda} g^\Lambda\right),
\end{equation}
where  ${\cal V} \equiv e^{K/2}\left(X^\Lambda(z), F_{\Lambda}(z) \right)$ are the covariantly holomorphic sections normalized as $1 ~=~ i\,\langle {\cal V}, \overline{\cal V}\rangle$.

The appropriate metric ansatz that captures static black holes solutions, interpolating between asymptotic $AdS_4$ space and near horizon $AdS_2\times S^2$ geometry, contains two warp factors. We parametrize it as
\begin{eqnarray}\label{metrica}
ds^{2}&=&-e^{2U(r)}dt^{2}+e^{-2U(r)}(dr^{2}+e^{2\psi(r)}d\Omega_{(2)}^{2})\ .
\end{eqnarray}
Upon this ansatz, the action  for the Lagrangian \eqref{StartingAction} reduces  (up to integration by part) to the form 
\begin{equation}\label{S1d}
	\begin{array}{rcl}
	S_{1d} &=&\displaystyle \int dr\left\{e^{2 \psi}\left[U'{}^2 - \psi'{}^2  + g_{i\bar \jmath}z^i{}' \bar z^{\bar \jmath}{}'+ e^{2U-4 \psi} V_{BH} + e^{-2U} V_g\right] - 1\right\}\\[5mm]
	&&\displaystyle + \int dr \frac{d}{dr}\left[e^{2 \psi}(2\psi'- U')\right].\\[5mm]
&=&\displaystyle \int dr\ \mc L^{eff}_{1d}\ + \int dr \frac{d}{dr}\left[e^{2 \psi}(2\psi'- U')\right]\ ,
	\end{array}
\end{equation}
the boundary contribution is exactly canceled by the Gibbons-Hawking boundary term.

One can explicitly check that is possible to rewrite (\ref{S1d}) as the sum of squares \cite{Dall'Agata:2010gj}
\begin{equation}\label{BPS2}
\begin{array}{rcl}
S_{1d}&=&\displaystyle\int dr\left\{-\frac12e^{2(U-\psi)}
{\cal E}^T\mathcal M {\cal E} -e^{2\psi}\left[ (\alpha'+\mathcal A_r) + 2 e^{-U} \, {\rm Re}(e^{-i \alpha} {\cal L}) \right]^{2}\right.\\[5mm]
&&\displaystyle-e^{2\psi}\left[\psi'-2e^{-U}\, {\rm Im}(e^{-i \alpha}\mathcal L) \right]^{2}-\left(1+\langle {\cal G},Q\rangle\right)\\[4mm]
&&\displaystyle\left.-2\frac{d\phantom{r}}{dr}\left[e^{2\psi-U}\,{\rm Im}(e^{-i \alpha}\mathcal L) +\,e^{U}\,{\rm Re}(e^{-i \alpha}\mathcal Z)\right]\right\},
\end{array}
\end{equation}
by introducing the symplectic vector
\begin{equation}
	{\cal E}^T \equiv  2e^{2 \psi}\left(e^{-U}{\mathrm{Im}}(e^{-i\alpha}{\cal V})\right)'\,^{T}-e^{2(\psi-U)}{\cal G}^{T}\Omega \mathcal M^{-1}+4e^{-U}(\alpha'+\mathcal A_r){\mathrm{Re}}(e^{-i\alpha}{\cal V})^{T}
	+Q^{T}.
\end{equation}
Equations of motions are then simply obtained by setting to zero the quantities that appear in each squared term.

\subsection{\label{rotation_charges}Rotation of charges: towards non-BPS solutions}

The equations of motion obtained in the previous subsections were shown to be equivalent to those obtained by the supersymmetry variations of the fermionic fields on the black hole solution \cite{Dall'Agata:2010gj}. However, the procedure of squaring the action \eqref{S1d} is not unique. 

In fact, we can act with a linear transformation on the black hole charges, and obtain a different set of first order equations with respect to those of \cite{Dall'Agata:2010gj}. The new solutions still satisfy the second order equations of motion of N=2 U(1)-gauged Supergravity, but now correspond to non supersymmetric configurations. In the following we are going to apply to the $U(1)$-gauged theory the procedure presented in \cite{Ceresole:2007wx}, to obtain a non-BPS flow.

Consider, in fact, a symplectic rotation acting on the Black Hole charges, given by a constant matrix $ S$, such that
\begin{eqnarray}\label{rotation-cond}
Q=(p^{\Lambda}\ ,q_{\Lambda})\rar \tilde Q\equiv S Q\ ,\qquad S^T\Omega S=\Omega\ , \qquad S^T\mc M S=\mc M\ ,
\end{eqnarray}
$S$ does not act on the scalars symplectic sections.  It has the same role and the properties \eqref{rotation-cond} are the same as those of the matrix $S$ introduced in Sec. 3 of \cite{Ceresole:2007wx}.  The black hole charges $Q$ enter the action \eqref{S1d} only through the black hole potential 
\begin{eqnarray}
V_{BH}&=&-\frac12 Q^T \mc M Q\ ,
\end{eqnarray}
which is left invariant by a matrix $S$ that obeys \eqref{rotation-cond}. Also the gauging potential $V_g$, that only depends on the scalars and FI terms, is left invariant, and so is the 1d action \eqref{S1d}.

We re-do the computation of Sec. \ref{BPS-squaring-L}, introducing a \textit{fake} central charge 
\begin{eqnarray}
\tilde{\mc Z}=\langle \tilde Q\,, \mc V\rangle\ ,
\end{eqnarray}
so that the 1d effective action (\ref{S1d}) can be squared as
\begin{equation}\label{non-BPS}
\begin{array}{rcl}
S_{1d}&=&\displaystyle\int dr\left\{-\frac12e^{2(U-\psi)}
{\tilde{\cal E}}^T\mathcal M {\tilde{\cal E}} -e^{2\psi}\left[ (\alpha'+\mathcal A_r) + 2 e^{-U} \, {\rm Re}(e^{-i \alpha} {\cal L}) \right]^{2}\right.\\[5mm]
&&\displaystyle-e^{2\psi}\left[\psi'-2e^{-U}\, {\rm Im}(e^{-i \alpha}\mathcal L) \right]^{2}-\left(1+\langle {\cal G},\tilde Q\rangle\right)\\[4mm]
&&\displaystyle\left.-2\frac{d\phantom{r}}{dr}\left[e^{2\psi-U}\,{\rm Im}(e^{-i \alpha}\mathcal L) +\,e^{U}\,{\rm Re}(e^{-i \alpha}\tilde{\mathcal Z})\right]\right\}\ ,
\end{array}
\end{equation}
where now
\begin{eqnarray}
\tilde{\mc E}^T \equiv  2e^{2 \psi}\left(e^{-U}{\mathrm{Im}}(e^{-i\alpha}{\cal V})\right)'\,^{T}-e^{2(\psi-U)}{\cal G}^{T}\Omega \mathcal M^{-1}+4e^{-U}(\alpha'+\mathcal A_r){\mathrm{Re}}(e^{-i\alpha}{\cal V})^{T}
	+\tilde Q^{T}\ .\nn\\
\end{eqnarray}
The first order equations are then
\begin{eqnarray}\label{nonBPSeom}
\tilde{\mc E}&=&0\ ,\nn\\
\psi'&=&2e^{-U}\, {\rm Im}(e^{-i \alpha}\mathcal L)\ ,\nn\\
(\alpha'+\mathcal A_r) &=&- 2 e^{-U} \, {\rm Re}(e^{-i \alpha} {\cal L}) \ ,
\end{eqnarray}
supplemented by the constraint
\begin{eqnarray}\label{nonBPSconstr}
1+\langle {\cal G},\tilde Q\rangle&=&0\ ,
\end{eqnarray}
and describe possibly extremal non-BPS black hole solutions.
In fact, the SUSY equations depend on the black hole charges $Q$ (defined, as usual, as the fluxes of the abelian gauge fields at infinity), while the flow equations above depend on the vector $\tilde Q$ .

Eventually, the black hole solution will differ from the BPS one because of the dependence of first order equations on $\tilde Q$ instead of $Q$. 

The flow equations for the scalar fields,  the warp factor $U$ and the constraint on the charges are affected by the symplectic rotation, while the equation of the warp factor $\psi$ remains the same as in the BPS case. Also the phase $\alpha$ satisfies an unchanged equation, but we recall that this phase is not an additional degree of freedom and its equation is implied by the others. 

The symplectic rotation is possible whenever a non-trivial matrix satisfying (\ref{rotation-cond}) exists. 
Let us notice that, analogously to the ungauged case, the choice $S = - \bfone$  gives the second branch of the BPS equations, obtained by a different choice of the phases for Killing spinor projectors. We refer to  Appendix \ref{2ndBPS-branch} for a more detailed discussion.

It is easy to find a matrix $S$ for the STU model with zero axions and all the moduli identified.
Indeed, in the case of the cubic $stu$ prepotential 
\begin{eqnarray}
F&=&\frac{X^1X^2X^3}{X^0}\ ,
\end{eqnarray}
if we identify all the moduli and look for the zero axions solutions
\begin{eqnarray}
s=t=u=-i\lambda\ ,
\end{eqnarray}
the most general matrix  satisfying (\ref{rotation-cond}) is 
\begin{eqnarray}
\left(
\begin{array}{cc|cc}
a&0&0&0\\0&A_{3\times3}&0&0\\ \hline
0&0&a&0\\0&0&0&A_{3\times3}
\end{array}
\right)\ ,\qquad a=\pm1\ ,\qquad A_{3\times3}^2=\bfone_{3\times3}\ ,
\end{eqnarray}
with $A_{3\times3}$ a 3x3 matrix.

\section{\label{3}The $t^3$ model}

Static regular black holes solutions in $\mc N=2$ $U(1)$-gauged Supergravity can be at most $1/4$-BPS, and have been derived in \cite{Cacciatori:2009iz,Dall'Agata:2010gj,kiril_stefan}. 
We are going to consider the $t^3$ model described by a cubic prepotential. After rewiewing the equations of motion and the BPS configuration, we derive a non-BPS solution by applying the rotation explained above to the supersymmetric configuration of charges $(p^0, q_1)$ and gauging $(g_0,g^1)$.

\subsection{\label{t3BPS} 1/4-BPS black hole solution for the  $t^3$ model}

Let us consider the prepotential 
\begin{eqnarray}\label{}
F&=&\frac{(X^{1})^3}{X^{0}}\ ,
\end{eqnarray}
which, by the choice of projective coordinates\ \, $t=X^{1}/X^{0}$,  becomes 
$F=t^3$. The symplectic sections are
\begin{eqnarray}\label{}
\mc V&=& \left(
L^{\Lambda}\ ,M_{\Lambda}\right)\ , \qquad \langle \mc V\ ,\bar{\mc V}\rangle=-i\ ,
\end{eqnarray}where
\begin{eqnarray}\label{}
L^{\Lambda}&=&e^{\mc K/2}\left(\begin{array}{c}
1\\t\end{array}\right)\ ,\qquad
M_{\Lambda}=e^{\mc K/2}\left(\begin{array}{c}
-t^3\\3t^2\end{array}\right)\ .
\end{eqnarray}
The K\"ahler potential is $\mc K=-\log(8\lambda^3)$,
which requires $\lambda>0$.

\subsubsection{\label{appsec1}The electric dyonic configuration}
We write the complex scalar field as $t=x-i\lambda$, and we consider a solution with no axions $x=0$, which is consistent with the choice of black hole and gauge charges $Q=(p^{0}, q_1)$, $\mc G=( g^{1},g_{0})$. The symplectic sections for our configuration are
\begin{eqnarray}\label{}
L^{\Lambda}&=&\frac1{2\sqrt2}
\left(\begin{array}{c}
1/\lambda^{3/2}\\
-i/\sqrt\lambda\end{array}\right)\ ,\qquad
M_{\Lambda}=\frac1{2\sqrt2}
\left(\begin{array}{c}
-i\lambda^{3/2}\\
-3\sqrt{\lambda}\end{array}\right)\ ,
\end{eqnarray}
The symplectic matrix describing the coupling of the scalars to the vectors is
\begin{eqnarray}\label{}
\mc N_{\Lambda\Sigma}&=&i\mc I_{\Lambda\Sigma}\ ,\qquad \mc I_{\Lambda\Sigma}=\left(\begin{array}{cc}
-\lambda^3& \\
&-3\lambda
\end{array}\right)\ ,
\end{eqnarray}
and the matrix $\mc M$ giving the black hole potential $V_{BH}$ is 
\begin{eqnarray}\label{}
\mc M&=&\left(
\begin{array}{cc}
\ \mc I\ \ & \\ \ \ & \mc I^{-1}
\end{array}
\right)\ ,
\end{eqnarray}
the central black hole and gauge charge are
\begin{eqnarray}\label{}
\mc Z&=&\langle Q,\mc V\rangle=\frac{i}{2\sqrt2}\left(
p^{0}\lambda^{3/2}-\frac{q_1}{\sqrt\lambda}
\right)\ ,\\
\mc L&=&\langle \mc G,\mc V\rangle=\frac{1}{2\sqrt2}\left(\frac{g_{0}}{\lambda^{3/2}}+3
 g^{1}\sqrt\lambda
\right)\ .
\end{eqnarray}
Knowing these central charges we can easily compute the phase $\alpha$ from
\begin{eqnarray}\label{}
e^{2i\alpha}&=&\frac{\mc Z-ie^{2A}\mc L}{\bar{\mc Z}+ie^{2A}\bar{\mc L}}\ ,
\end{eqnarray}
and thus $\alpha=\pm\pi/2$.
The requirement of asymptotic $AdS_{4}$ space is $D_ t\mc L|_{\infty}=0$, which fixes the value of the scalar at infinity to be
\begin{eqnarray}\label{}
\lambda_{\infty}=\sqrt{\frac{g_{0}}{ g^{1}}}\ ,
\end{eqnarray}
that requires $g_0\cdot g^1>0$.

\subsubsection{\label{appsec2}Equations of motion}
The BPS equations relative to this configuration, as derived in \cite{Dall'Agata:2010gj} from the BPS squaring \eqref{BPS2}, are
\begin{eqnarray}\label{BPSeq}
2e^{2\psi}\left( e^{-U}\re\mc V \right)'+e^{2(\psi-U)}\Omega\mc M\mc G+Q&=&0\ ,\nn\\
(e^{\psi})'=2 e^{\psi-U}\re\mc L\ .
\end{eqnarray}
Explicitely, one has
\begin{eqnarray}\label{}
\Omega\mc M\mc G&=&\left(\begin{array}{c}
g_{0}/\lambda^3\\
0\\0\\
-3 g^{1}{\lambda}\end{array}
\right)=8
\left(\begin{array}{c}
g_{0}(L^{0})^{2}\\
0\\0\\
-\frac13 g^{1}(M_{1})^{2}\end{array}
\right)\ ,
\end{eqnarray}
so if we define four positive functions
\begin{eqnarray}\label{defH}
H^{0}&=&L^{0}e^{-U}\ ,\quad H_{i}=-\frac13M_{i}e^{-U}\ , 
\end{eqnarray}
we can rewrite (\ref{BPSeq}) as 
\begin{eqnarray}\label{BPSH}
2e^{2\psi}\left(\begin{array}{c}
\p_{r}H^{0}+4 g_{0}(H^{0})^{2}\\
-3\p_{r}H_{1}-12  g^{1}(H_{1})^{2}\end{array}
\right)=
\left(
\begin{array}{c}
-p^{0}\\ -q_{1} 
\end{array}
\right)\ ,\qquad \psi'=2(g_{0}H^{0}+3 g^{1}H_{1})\ .
\end{eqnarray}
Notice that we have
\begin{eqnarray}
e^{-2U}&=&8\sqrt{H^0(H_1)^3}\ ,\qquad\quad \lambda=\sqrt{\frac{H_1}{H^0}}\ .
\end{eqnarray}
Following the assumptions of \cite{Cacciatori:2009iz} we make the following ansatz
\begin{eqnarray}\label{ans}
H^{0}&=&e^{-\psi}(\alpha^{0}r+\beta^{0})\ ,\quad H_{1}=e^{-\psi}(\alpha_{1}r+\beta_{1})\ ,\quad 
\psi=\log(a r^{2}+c)\ ,
\end{eqnarray}
and we look for $-c=r_{h}^{2}$, so that $\psi=\log(a\,r^{2}-r_{h}^{2})$.
The equations (\ref{BPSH}) now become algebraic equations
\begin{eqnarray}\label{alg1}
\frac{p^{0}}2&=&\alpha^{0}r_{h}^{2}-4g_{0}(\beta^{0})^{2} \qquad 
\alpha^{0}=\frac{a}{4g_{0}}\ ,\nn\\
\frac{q_{1}}6&=&-\alpha_{1} r_{h}^{2}+4 g^{1}(\beta_{1})^{2} \qquad 
\alpha_{1}=\frac a{4 g^{1}}\quad\ ,\nn\\
&&g_{0}\beta^{0}+3 g^{1}\beta_{1}=0\ ,
\end{eqnarray}
that have to be supplemented by the BPS constraint \cite{Dall'Agata:2010gj}
\begin{eqnarray}\label{constr}
\langle \mc G,Q\rangle=g_{0}p^{0}- g^{1}q_{1}=-1\ .
\end{eqnarray}
Without loss of generality one can restrict to $a=1$ (as explained in \cite{Cacciatori:2009iz}); from (\ref{alg1}) e (\ref{constr}) we are left then with the system of 4 equations
\begin{eqnarray}\label{}
p^{0}&=&\frac{r_{h}^{2}}{2g_{0}}-8g_{0}(\beta^{0})^{2} \qquad
\frac{q_1}3=-\frac{r_{h}^{2}}{2 g^1}+8 g^1(\beta_1)^{2}\nn\\
0&=&g_{0}\beta^{0}+3 g^1 \beta_1\ \quad  \qquad g_{0}p^{0}- g^1 q_1=-1
\end{eqnarray}
and 7 unknowns $\{q_1,p^{0}, g^1,g^{0},\beta_1,\beta^{0}, r_{h}\}$; we choose to parametrize the solution with $q_1, g^1$ and $g^{0}$. Moreover, we see that if we define the hatted quantities
\begin{eqnarray}\label{}
\hat q_1\equiv q_1\cdot g^1 \qquad \hat p^{0}\equiv p^{0}\cdot g_{0} \qquad \hat\beta_1\equiv \beta_1\cdot g^1 \qquad \hat\beta^{0}\equiv\beta^{0}\cdot g_{0}\ ,
\end{eqnarray} 
and choose $g^{0}>0$, $ g^1>0$, the equations become simply
\begin{eqnarray}\label{}
\hat p^{0}&=&\frac{r_{h}^{2}}{2}-8(\hat\beta^{0})^{2} \qquad
\frac{\hat q_1}3=-\frac{r_{h}^{2}}{2}+8(\hat\beta_1)^{2}\nn\\
0&=&\hat\beta^{0}+3\hat\beta_1\ \quad  \qquad \hat p^{0}-\hat q_1=-1
\end{eqnarray}
We parametrize the solution of these system with $\hat q_1$; we then have
\begin{eqnarray}\label{}
\hat p^{0}=\hat q_1-1\qquad \hat\beta_1=-\frac{\sqrt{1-4\hat q_1/3\,}}{8} \qquad \hat\beta^{0}=\frac38\sqrt{1-4\hat q_1/3\,}\qquad r_{h}=\frac{\sqrt{1-4\hat q_1\,}}{2}
\end{eqnarray}
in fact one can show that a regular solution with all positive gauge charges cannot have $\beta_1>0$. We also have to check that the functions in (\ref{defH}) are well defined, in particular that they are positive throughout the flow; this imply, given $\hat\beta_1<0$, that $r_{h}>-2\hat\beta_1$ which results in
\begin{eqnarray}\label{}
\hat q_1<0\quad \Rightarrow \quad q_1<0 \ \ \cup \ \ p^{0}<0\ .
\end{eqnarray}

To summarize, we have a black hole solution whose scalars and metric warp factors are parametrized by the functions in (\ref{defH}, \ref{ans}), with $\alpha^{0}$, $\alpha_{i}$ given in (\ref{alg1}) and 
\begin{eqnarray}\label{recap}
p^{0}=\frac{ g^1\, q_1-1}{g_{0}}\qquad \beta=-\frac{\sqrt{1-4\, g^1\, q_1/3\,}}{8 g^1} \qquad \beta^{0}=\frac3{8g^{0}}\sqrt{1-4\, g^1\, q_1/3\,}\qquad r_{h}=\frac{\sqrt{1-4\, g^1\, q_1\,}}{2}\nn\\
\end{eqnarray}
we are left with the freedom to choose $q_1<0$, $g_{0}>0$ and $g^1>0$. 
The scalar is
\begin{eqnarray}\label{}
\lambda&=&\sqrt{\frac{H_{1}}{H^{0}}}=\lambda_{\infty}\sqrt{\frac{2r-\sqrt{1-4\, g^1\, q_1/3}}{2r+3\sqrt{1-4\, g^1\, q_1/3}}}\ ,
\end{eqnarray}
 and, at the horizon, it takes the value
\begin{eqnarray}\label{}
\lambda_{h}&=&{\lambda_{\infty}\over \sqrt2} \sqrt{\frac{-1+2\, g^1 q_1+\sqrt{1-16\, g^1\, q_1/3+48\,( g^1)^{2}(q_1/3)^{2}}}{1-\, g^1\,q_1}}\ .
\end{eqnarray}
The asymptotically $AdS_{4}$ metric, solution of the $STU$-model in $U(1)$-gauged $\mc N=2$ supergravity with $AdS_{2}\times S^{2}$ horizon is 
\begin{eqnarray}\label{}
ds^{2}&=&-e^{2U}dt^{2}+e^{-2U}dr^{2}+e^{-2U+2\psi}(d\theta^{2}+\sin\theta\,^{2}\,d\phi^{2})\ ,
\end{eqnarray}
where the warp factors are
\begin{eqnarray}\label{}
e^{2\psi(r)}&=&(r^{2}-r_{h}^{2})^{2}\ ,\nn\\
 e^{2U}&=&
\frac{2\sqrt{g^{0}( g^1)^{3}}(r^{2}-r_{h}^{2})^{2}}{\left(
r-\frac12\sqrt{1-4\, g^1\,q_1/3}
\right)^{3/2}\left(
r+\frac32\sqrt{1-4\, g^1\,q_1/3}
\right)^{1/2}}\nn\\
&=&\frac{2\sqrt{g^{0}( g^1)^{3}}(r^{2}-r_{h}^{2})^{2}}{\left(
r-\sqrt{r_{h}^{2}+2\, g^1\,q_1/3}
\right)^{3/2}\left(
r+3\sqrt{r_{h}^{2}+2\, g^1\,q_1/3}
\right)^{1/2}}\ ,
\end{eqnarray}
where we recall that, by definitions of $H$\rq{}s, $e^{-2U(r)}=8\sqrt{H_{0}H_{1}^3}$.
The entropy is given by the warp factor $e^{2A}|_{h}=e^{2\psi(r_{h})-2U(r_{h})}=S$, 
\begin{eqnarray}\label{}
e^{2A(r_{h})}&=&\frac1{2\sqrt{g^{0}( g^1)^{3}}}{\left(
r_h-\sqrt{r_{h}^{2}+2\, g^1\,q_1/3}
\right)^{3/2}\left(
r_h+3\sqrt{r_{h}^{2}+2\, g^1\,q_1/3}
\right)^{1/2}}=\nn\\
&=&\frac1{8\sqrt{g^{0}( g^1)^{3}}}\left(
\sqrt{1-4 g^1 q_1}-\sqrt{1-4 g^1 q_1/3}
\right)^{3/2}\sqrt{
\sqrt{1-4 g^1 q_1}+3\sqrt{1-4 g^1 q_1/3}}\ .\nn\\
\end{eqnarray}

\subsection{\label{non-BPS-t3-extremal}The extremal non-BPS $t^3$ configuration}

We can now exploit the trick derived in Sec. \ref{trick} to derive the non-BPS solution from the dyonic supersymmetric one. In particular, we present here the case where the matrix $S$ satisfying \eqref{rotation-cond} is 
\begin{eqnarray}
S=\left(\begin{array}{cc} \textbf{1} &\ \ 0\\0&-\textbf{1}\end{array}\right)\ .
\end{eqnarray}

The rotated equations of motion for the non-BPS flow are, from \eqref{nonBPSeom},
\begin{eqnarray}
2e^{2\psi}\left(\begin{array}{c}
\p_{r}H^{0}+4 g_{0}(H^{0})^{2}\\
-3\p_{r}H_{1}-12  g^{1}(H_{1})^{2}\end{array}
\right)=
\left(
\begin{array}{c}
-\tilde p^{0}\\ -\tilde q_{1} 
\end{array}
\right)\ ,\qquad \psi'=g_{0}H^{0}+3 g^{1}H_{1}\ .
\end{eqnarray}
The charges further satisfy the constraint  \eqref{nonBPSconstr}
\begin{eqnarray}\label{}
\langle \mc G,\tilde Q\rangle&=&g_{0}\tilde p^{0}- g^{1}\tilde q_{1}=-1 \qquad \Rightarrow \qquad
g_{0} p^{0}+ g^{1}q_{1}=-1\ .
\end{eqnarray}
We restrict to the zero axions case: $t=-i\lambda$.
We still consider a flow that starts at asymptotic infinity from the supersymmetric $AdS_4$ space, where the scalar has the value
\begin{eqnarray}
\lambda_\infty&=&\sqrt{\frac{g_0}{ g^1}}\ .
\end{eqnarray}
The solution is then given in terms of the following ansatz
\begin{eqnarray}\label{}
H^{0}&=&e^{-\psi}(\alpha^{0}r+\beta^{0})\ ,\quad H_{1}=e^{-\psi}(\alpha_{1}r+\beta_{1})\ ,\quad 
\psi=\log(r^{2}-r_h^2)\ ,
\end{eqnarray} 
by the choice of parameters 
\begin{eqnarray}\label{}
\tilde p^{0}=\frac{\, g^1\, \tilde q_1-1}{g_{0}}\qquad \beta_1=-\frac{\sqrt{1-4\, g^1\, \tilde q_1/3\,}}{8 g^1} \qquad \beta^{0}=\frac3{8g^{0}}\sqrt{1-4\, g^1\, \tilde q_1/3\,}\qquad r_{h}=\frac{\sqrt{1-4\, g^1\,\tilde q_1\,}}{2}\ ,\nn
\end{eqnarray}
where $\tilde q_1<0$, $g_{0}>0$ and $ g^1>0$. This means that, in terms of the black hole charges, the non-BPS extremal solution is given by
\begin{eqnarray}\label{}
p^{0}=-\frac{1+\, g^1\, q_1}{g_{0}}\qquad \beta_1=-\frac{\sqrt{1+4\, g^1\, q_1/3\,}}{8 g^1} \qquad \beta^{0}=\frac3{8g^{0}}\sqrt{1+4\, g^1\, q_1/3\,}\qquad r_{h}=\frac{\sqrt{1+4\, g^1\, q_1\,}}{2}\ . \nn
\end{eqnarray}
Here we have the freedom to choose $q_1>0$, $g_{0}>0$ and $ g^1>0$.

The field profile is given by
\begin{eqnarray}\label{}
\lambda&=&\lambda_{\infty}\sqrt{\frac{2r-\sqrt{1+4\, g^1\, q_1/3}}{2r+3\sqrt{1+4\, g^1\, q_1/3}}}\ ,
\end{eqnarray}
the metric solution with spherical horizon and non supersymmetric first order flow is 
\begin{eqnarray}\label{}
ds^{2}&=&-e^{2U}dt^{2}+e^{-2U}dr^{2}+e^{-2U+2\psi}(d\theta^{2}+\sin\theta\,^{2}\,d\phi^{2})\ ,
\end{eqnarray}
with warp factors 
\begin{eqnarray}\label{}
e^{2\psi(r)}&=&(r^{2}-r_{h}^{2})^{2}\ ,\nn\\
 e^{2U}&=&
\frac{2\sqrt{g^{0}( g^1)^{3}}(r^{2}-r_{h}^{2})^{2}}{\left(
r-\frac12\sqrt{1+4\, g^1\,q_1/3}
\right)^{3/2}\left(
r+\frac32\sqrt{1+4\, g^1\,q_1/3}
\right)^{1/2}}\nn\\
&=&\frac{2\sqrt{g^{0}( g^1)^{3}}(r^{2}-r_{h}^{2})^{2}}{\left(
r-\sqrt{r_{h}^{2}-2\, g^1\,q_1/3}
\right)^{3/2}\left(
r+3\sqrt{r_{h}^{2}-2\, g^1\,q_1/3}
\right)^{1/2}}\ ,
\end{eqnarray}
where we recall that, by definitions of $H$\rq{}s, $e^{-2U(r)}=8\sqrt{H_{0}H_{1}^3}$.
The entropy is given by the warp factor $e^{2A}|_{h}=e^{2\psi(r_{h})-2U(r_{h})}$, 
\begin{eqnarray}\label{}
e^{2A(r_{h})}&=&\frac1{2\sqrt{g^{0}( g^1)^{3}}}{\left(
r_h-\sqrt{r_{h}^{2}-2\, g^1\,q_1/3}
\right)^{3/2}\left(
r_h+3\sqrt{r_{h}^{2}-2\, g^1\,q_1/3}
\right)^{1/2}}=\nn\\
&=&\frac1{8\sqrt{g^{0}( g^1)^{3}}}\left(
\sqrt{1+4 g^1 q_1}-\sqrt{1+4 g^1 q_1/3}
\right)^{3/2}\sqrt{
\sqrt{1+4 g^1 q_1}+3\sqrt{1+4 g^1 q_1/3}}\ .\nn\\
\end{eqnarray}

\subsubsection{\label{dir_nonBPS-tcube}Dirac quantization condition}

For a globally consistent interacting theory the gravitinos and black hole charges have to satisfy the Dirac quantization condition. This requirement arises from the fact that, in the abelian $\mathcal{N}=2$ gauged supergravity model taken into consideration, the FI parameters $g_{\Lambda}$ and $g^{\Lambda}$ are respectively the electric and magnetic charges of the gravitinos \cite{romans} \cite{kiril_stefan}.

The BPS solutions found so far obey the Dirac-quantization constraint from eq. \eqref{dirac_ultima} ($\hbar =1$) 
\begin{equation}\label{dirac}
g_{0} p^{0} - g^1 q_1= n \qquad n \in \mathbb{Z}\,,
\end{equation}
where supersymmetry fixes the number $n$ to be $\pm 1$ \cite{Dall'Agata:2010gj,kiril_stefan}. The non-BPS solutions, in general, do not satisfy the Dirac quantization condition, but they do satisfy, by construction, another relation. Let us for instance focus on the solution obtained by changing the sign of $q_1$. It satisfies
\begin{equation}
g_{0} p^{0} + g^1 q_1= -1\,.
\end{equation}
The charges, in this case, written in function of the parameter $\beta_1$, are
\begin{equation}
p^0=-\frac{1}{g_0} \left(\frac14 +48 (\beta_1 {g^1})^2 \right)\,,  \qquad q_1= - \frac{1}{ {g^1}}  \left(\frac34 -48 (\beta_1 {g^1})^2 \right)\,.
\end{equation}
With these values of charges, using the relation \eqref{recap}, we have:
\begin{equation}\label{rhfunzm}
g_{0} p^{0} - g^1 q_1= \frac12 -96 (g^1 \beta_1)^2= -2 r_h^2 - \frac12\,.
\end{equation}
It turns out that the value of $r_h$, that determines the radial position of the horizon, enters in the quantization condition, and is constrained to satisfy 
\begin{equation}
-\frac12 -2r_h^2 \in \mathbb{Z} \,.
\end{equation}
Whenever $r_h$ fulfills this condition, we are able to build a tower of states of extremal black holes with more generic $n$ integer.  We recall that, in order to have proper black holes (finite nonzero area of the event horizon) the parameter $r_h$ has to be positive. 

In addition to it, this requirement sets some constraints on the charges in the configuration. In particular, in this example we have:
\begin{equation}
g_{0} p^{0} + g^1 q_1= -1\,,
\end{equation}
\begin{equation}
g_{0} p^{0} - g^1 q_1= n\,.
\end{equation}
That gives
\begin{equation}\label{quant-cond-a1}
2 g_0 p^0=-1+n \qquad 2 g^1 q_1=-(1+n)\,.
\end{equation}
This restricts the lattice of possible charges, and, 
in particular, only values of $n=-m$, $m~\in~\bb N~\backslash~ \{0,1\}$ are allowed.
Together, the quantization conditions \eqref{quant-cond-a1} fix the solution in terms of the electric-magnetic charges plus the quantization integer parameter $m$. The non-BPS black hole solution for the scalar field of Sec. \ref{non-BPS-t3-extremal} can be expressed as
\begin{eqnarray}
\lambda&=&\lambda_{\infty}\sqrt{\frac{2r-\sqrt{1+2(m-1)\,/3}}{2r+3\sqrt{1+2\, (m-1)/3}}}\ ,
\end{eqnarray}
parametrized by this integer $m$. The warp factors of the metric ansatz (\ref{metrica}) are
\begin{eqnarray}\label{}
e^{2\psi(r)}&=&(r^{2}-r_{h}^{2})^{2}\ ,\nn\\
 e^{2U}&=&\frac{2\sqrt{g_{0}( g^1)^{3}}(r^{2}-r_{h}^{2})^{2}}{\left(
r-\sqrt{r_{h}^{2}-\, (m-1)/3}
\right)^{3/2}\left(
r+3\sqrt{r_{h}^{2}-(m-1)/3}
\right)^{1/2}}\ ,
\end{eqnarray}
with, from \eqref{rhfunzm},  
$r_h=\frac{\sqrt{2m-1}}2
$. The entropy is
\begin{eqnarray}\label{}
e^{2A(r_{h})}&=&\frac1{2\sqrt{g_{0}( g^1)^{3}}}{\left(
r_h-\sqrt{r_{h}^{2}-(m-1)/3}
\right)^{3/2}\left(
r_h+3\sqrt{r_{h}^{2}-(m-1)/3}
\right)^{1/2}}=\nn\\
&=&\frac{2\sqrt{|p^0|q_1^3}}{\sqrt{m^2-1}(m-1)}{\left(
r_h-\sqrt{r_{h}^{2}-(m-1)/3}
\right)^{3/2}\left(
r_h+3\sqrt{r_{h}^{2}-(m-1)/3}
\right)^{1/2}}\ .\nn\\
\end{eqnarray}
Notice that the entropy is given by an expression which is nothing but the entropy of the black hole in the corresponding ungauged Supergravity configuration, corrected by a factor that only depends on the quantization parameter $m$. We can write, more explicitely,
\begin{eqnarray}
e^{2A(r_{h})}&=&\frac12\frac{\sqrt{ |p^0|q_1^3}}{(m-1)\sqrt{m^2-1}}
\left(\sqrt{2m-1}-\frac1{\sqrt3}\sqrt{2m+1}
\right)^{3/2}
\left(\sqrt{2m-1}+{\sqrt3}\sqrt{2m+1}
\right)^{1/2}
\ .\nn\\
\end{eqnarray}
The quantity $\sqrt{ |p^0|q_1^3}$ corresponds to the quartic invariant of the duality group of the theory. It could be interesting to analyze further the duality properties for the gauged solutions, in analogy with the ungauged case.

\subsection{\label{nonextr_tcube}Nonextremal generalization of the $t^3$ solution}\label{nonextremal}

We now turn to the nonextremal generalization, following the general procedure introduced in \cite{ours_nonextr}. We choose the  ansatz for the function $\psi$ as
\begin{equation}\label{nonextr_new_sol}
e^{2 \psi} = r^2 f(r) = r^2  \left( r^2+c - \frac{\mu}{r}+\frac{Q}{r^2} \right)\ ,
\end{equation}
and the warp factor
\begin{equation}
e^{2U} = e^{\mathcal{K}} f(r)\ .
\end{equation}
Furthermore, $e^{2A(r)} = r^2 e^{-\mathcal{K}}$. We keep the same form of the sections as in the BPS case, namely \eqref{ans}.

This guess for the form of the nonextremal solution is then followed by brute-force solving the Einstein's equations of motion, namely (2.16-18) of \cite{ours_nonextr}.
 It turns out that the equations of motion are satisfied if the parameters present in \eqref{ans} assume the values \footnote{For consistency with the parametrization in \cite{ours_nonextr}, and since $\beta_1$ is taken to be the nonextremality parameter, we drop the subscript and from now on we simply intend $\beta\equiv\beta_1$.}
\begin{equation}
\alpha_0=\frac{1}{4g_0} \qquad \alpha^1= \frac{1}{4g^1} \qquad \beta_0=-\frac{3 g^1 \beta}{g_0}\ ,
\end{equation}
as in the BPS case, see \eqref{alg1} . Furthermore, the remaining parameters that determine the warp factors are
\begin{equation}\label{c1}
c= 1 - 96 \beta^2 (g^1)^2\,,
\end{equation}
\begin{equation}\label{mu}
\mu= 8 \beta g_1 + 512 \beta^3 (g^1)^3 - \frac{g_0^2 (p^0)^2}{ 4 \beta g^1} + \frac{g^1 q_1^2}{36 \beta}\,,
\end{equation}
\begin{equation}\label{c2}
Q= -48 \beta^2 (g^1)^2 - 768 \beta^4 (g^1)^4 + (g_0)^2 (p^0)^2 +  \frac{ (g^1)^2 q_1^2}{3}\,.
\end{equation}
For the moment we have left the charges $p^0, q_1$ unconstrained. 
The function $e^{\mathcal{K}}$ assumes the form
\begin{equation}
e^{\mathcal{K}}=\frac{2 \sqrt{g_0 (g^1)^3}}{\left(r+4 \beta g^1\right)^{3/2}\left(r-12 \beta g^1\right)^{1/2}}\,.
\end{equation}
The functional dependence resembles the one of the BPS case. However, in the nonextremal solution the parameter $\beta$ and the charges $q_1$ and $p^0$ are not related to each other; they are three independent quantities, among which $\beta$ plays the role of the nonextremality parameter.
The singularities $r_{s,1}= - 4 \beta g^1$ and $r_{s,2}=12 \beta g^1$ are the points in which $e^{\mathcal{K}}$ blows up, and one can check that for a suitable range of parameters there is a horizon shielding them.

In order to have a physical solution we need to impose the Dirac quantization condition \eqref{dirac}.
If we want a deformation over the BPS state described in the previous section, we should impose one of the following relation between the charges
\begin{equation}\label{quant}
g_0 p^0-g^1 q_1=\pm1\,,
\end{equation}
so that the state preserves asymptotically some supersymmetry.  The parameters $\mu$ and $Q$ then depend just on the $q_1$ parameter, having eliminated the dependence on $p^0$ through \eqref{quant}.
\begin{equation}\label{newparam1}
\mu=-\frac{1}{4 \beta g^1} + 8 \beta g^1 + 512 \beta^3 (g^1)^3 \mp \frac{ q_1}{ 2\beta} - \frac{2 g^1 q_1^2}{9\beta}\,,
\end{equation}
\begin{equation}\label{newparam2}
Q= 1 - 48 \beta^2 (g^1)^2 - 768 \beta^4 (g^1)^4 \pm 2 g^1 q_1 + \frac43 (g^1)^2 q_1^2\,,
\end{equation}
One can verify that the solution above has a finite nonzero area of the event horizon for a suitable choice of parameters.

\subsection{Product of the areas}

In this section we compute the product of the areas of the horizons for the new dyonic nonextremal solution we found in the previous section. It is true in a lot of examples \cite{Cvetic:2010mn,Galli:2011fq,Castro:2012av}, that for nonextremal black hole solutions the product between the areas of the inner and outer event horizons does not depend on the mass. In particular, such product depends just on the quantized electric and magnetic charges. This fact might be a hint for some underlying microscopic structure  \cite{Larsen:1997ge}. 

For AdS black holes the result holds if we take the product of the square of the four roots of the $g_{tt}$ component of the metric \cite{Cvetic:2010mn}. In the case of the nonextremal solution of Section \ref{nonextr_tcube} we have:
\begin{equation}\label{prodkappa}
 \prod_{{\alpha}=1}^4 Area_{\alpha} = (4 \pi)^4 \prod_{\alpha=1}^4 e^{2A(r_{\alpha})}=(4 \pi)^4 \prod_{\alpha=1}^4 e^{-\mathcal{K}(r_{\alpha})} r_{\alpha}^2 \,,
\end{equation} 
where the function $e^{2A(r)}$ is of the form
\begin{equation}\label{acca}
e^{2A(r)}= {\rm const} \times \sqrt{(r-r_{s,1})(r-r_{s,2})^3}\,,
\end{equation}
with $r_{s,1/2}$ the location of the singularities. Following Section 6 of \cite{ours_nonextr},
\begin{equation}
e^{2U(r)}=e^{\mathcal{K}} \left(r^2 +c- \frac{\mu}{r} +\frac{Q}{r^2} \right)=\frac{e^{\mathcal{K}}}{r^2} \left(r^4 +c r^2- \mu r +Q \right) = \frac{e^{\mathcal{K}}}{r^2} \prod_{\alpha=1}^4 (r-r_{\alpha})\,.
\end{equation}
The coefficient of lowest degree in $r$, namely $Q$, gives the value the product of all the roots $r_1 r_2 r_3 r_4$.
We now first make the redefinition
\begin{equation}
r'=r-r_{s,1}\,,
\end{equation}
and we express the warp factor in terms of $r'$. In a similar way the coefficient of lowest degree in $r'$, from now on denoted by $\kappa_1$, represents the product of all the $r'$ roots: $r'_1 r'_2 r'_3 r'_4$. This coefficient turns of to be:
\begin{equation}\label{kappa-magn}
\kappa_1= r_{s,1}^4 +c r_{s,1}^2-\mu  r_{s,1}+Q= (2 g^1 q_1/3)^2\,,
\end{equation}
where the values of $c$, $\mu$ and $Q$ are given respectively in \eqref{c1}, \eqref{mu} and \eqref{c2}. Repeating the procedure for $r_{s,2}$ gives $\kappa_2=( 2 g_0 p^0)^2$, so that we have what we need to compute the area product. Using \eqref{acca} and \eqref{prodkappa}, we arrive at
\begin{equation}\label{prod1}
 \prod_{\alpha=1}^4 A_{\alpha}= ({\rm const})^4 (4 \pi)^4 \sqrt{ \kappa_1 \kappa_2^3}= \frac{(4 \pi)^4}{27} \frac{p^0 (q_1)^3}{g_0 (g^1)^3}\,.
\end{equation}
We see then that the result  depends only on the black hole and gravitino charges. Furthermore, if we impose the relations \eqref{quant-cond-a1} (with $m=-n$) between the gravitino charges and the black hole charges, the area product assumes this form:
\begin{equation}\label{prod_con_quant}
 \prod_{\alpha=1}^4 A_{\alpha}= (4 \pi)^4 \frac{1}{27\times16} \frac{(m+1)^2(m^2-1)}{g_0^2 (g^1)^6}\,.
\end{equation}

\section{\label{4}The $F=-2i\sqrt{X^0(X^1)^3}$ model}

In this section we consider regular solutions of the model with prepotential $F=-2i\sqrt{X^0(X^1)^3} $, in presence of gauging charges $\mathcal{G}=(g_0, g_1)$. These solution are purely magnetic: $q_0=q_1=0$. We will first describe the $1/4$ BPS solution found by \cite{Cacciatori:2009iz} and \cite{kiril_stefan}, we then discuss the non-BPS solution generated by the procedure explained in Section \ref{rotation_charges}. Finally we review the nonextremal generalization of the magnetic solution, already found in \cite{ours_nonextr}, and we comment on the product of the four areas. In Appendix \ref{EquivPrep} we show the equivalence of this configuration to the dyonic one presented in the previous section.

\subsection{\label{squareroot_BPS}The magnetic BPS configuration}

In this subsection we describe the general setting in presence of the prepotential  $F=-2i\sqrt{X^0(X^1)^3}$. The gauging charges are $\mathcal{G}=(g_0, g_1)$ while the black hole charges are $Q=(p^0,p^1)$. The symplectic sections
 are
\begin{eqnarray}\label{}
\mc V&=& \left(
L^{\Lambda}\ ,M_{\Lambda}\right)\,,
\end{eqnarray}where
\begin{eqnarray}\label{}
L^{\Lambda}&=&e^{\mc K/2}\left(\begin{array}{c}
X^0 \\ X^1\end{array}\right)\ ,\qquad
M_{\Lambda}=e^{\mc K/2}\left(\begin{array}{c}
-i \sqrt{\frac{(X^1)^3}{X^0}}\\-3i \sqrt{X^1 X^0} \end{array}\right)\,.
\end{eqnarray}
The K\"ahler potential is 
\begin{equation}
\mathcal{K}=-log[i(\overline{X}^{\Lambda}F_{\Lambda} - \overline{F}_{\Lambda}X^{\Lambda})]=-log[X^0 \overline{X}^0 (\sqrt{z}+\sqrt{\bar{z}})^3]\,.
\end{equation}
We consider a solution with real and positive scalar $z=\frac{X^1}{X^0}$. The period matrix is then purely imaginary
\begin{eqnarray}\label{}
\mc N_{\Lambda\Sigma}&=&i\mc I_{\Lambda\Sigma}\ ,\qquad \mc I_{\Lambda\Sigma}=\left(\begin{array}{cc}
-z^{3/2}& \\
&-3\sqrt{\frac{1}{z}}
\end{array}\right)\ ,
\end{eqnarray}
and the matrix $\mc M$ is 
\begin{eqnarray}\label{}
\mc M&=&\left(
\begin{array}{cc}
\ \mc I\ \ & \\ \ \ & \mc I^{-1}
\end{array}
\right)\,.
\end{eqnarray}
The scalar potential is then
\begin{equation}
V_g(z, \overline{z})= -\left(\frac{g_0 g_1}{\sqrt{z}}+ \frac{g_1^2}{3} \sqrt{z}\right)\,,
\end{equation}
and the asymptotic value of the scalar field, for which the scalar potential is extremized, is
\begin{equation}\label{}
z_{\infty}=\sqrt{\frac{3g_{0}}{ g_{1}}}\,.
\end{equation}
This gives $V_g(z_{\infty}, \bar{z}_{\infty}) <0$, so that the solution asymptotes to AdS$_4$.

\subsubsection{1/4-BPS solution}

The $1/4$ BPS purely magnetic ($q_{0}=q_1=0$) solutions found in \cite{Cacciatori:2009iz} and \cite{kiril_stefan} are described by these warp factors: 
\begin{equation}\label{e2U}
e^{2U(r)}= e^{\mathcal{K}}\frac{\left(r^2 -r_h^2 \right)^2}{r^2}\,, \qquad
e^{A(r)}= r e^{-\mathcal{K}/2}\,.
\end{equation}
Furthermore, the BPS solutions satisfy 
\begin{equation}\label{quant_ours}
g_{\Lambda}p^{\Lambda}= \pm 1 \, ,
\end{equation}
consistently with the choice of the BPS branch under consideration (see Appendix A for details).
From here we see that supersymmetry constrains the possible allowed value of the Dirac quantization relation, namely, as already mentioned, just the values $n=\pm 1$ of \eqref{dirac_ultima} are possible. 
 
For simplicity, let us focus on the branch of solutions that satisfy $g_{\Lambda} p^{\Lambda}= -1$, the others can be treated in full similarity.
 The sections are harmonic functions and $z$ is:
\begin{equation}\label{param_x}
 X_0 = \alpha_0 + \frac{\beta_0}{r}\,, \qquad X_1 = \alpha_1 + \frac{\beta_1}{r}\,, \qquad z= \frac{X_1}{X_0}\,.
\end{equation}
The parameters appearing in \eqref{param_x} are constrained by the Killing spinor equations (we are dealing here with the solutions found in \cite{kiril_stefan}, where $\alpha$, phase of the Killing spinor, is $\alpha=0
$) to be
\begin{equation}\label{param_BPS}
\alpha_0= - \frac{1}{4 g_0}\,, \qquad \beta_0= -\frac{\xi_1 \beta_1}{\xi_0}\,, \qquad \alpha_1= - \frac{3}{4 g_1}\,, \qquad
r_h^2=\frac{16}{3}(g_1 \beta_1)^2-\frac12\,.
\end{equation}
Furthermore, the value of $\beta_1$ is fixed in terms of the magnetic charges. One can also eliminate $p^0$ thanks to \eqref{quant_ours}, so that:
\begin{equation}
\beta_1=-\frac{3 \sqrt{ 1+4(p^1 g_1)/3 }}{8 g_1}\,, \qquad r_h= \frac{\sqrt{1+4g_1 p^1} }{2}\,.
\end{equation}
Vice versa, the magnetic charges can be expressed in terms of $\beta_1$:
\begin{equation}
p^0=  - \frac{2}{g_0} \left(\frac18 +\frac{8 (g_1 \beta_1)^2}{3} \right)\,,\qquad
p^1= -\frac{2}{g_1} \left(\frac38 -\frac{8 (g_1 \beta_1)^2}{3} \right)\,.
\end{equation}
The warp factor assumes this form: 
\begin{eqnarray}\label{}
 e^{2U}&=&
\frac{2\sqrt{g_{0}( g_1)^{3}}(r^{2}-r_{h}^{2})^{2}}
{\left(r+ \frac{3}{2} \sqrt{ 1+4(p^1 g_1)/3 }\right)^{1/2}
\left(3r-\frac32 \sqrt{ 1+4(p^1 g_1)/3 }\right)^{3/2}} \,,
\nn\\
\end{eqnarray}
and the entropy is
\begin{equation}\nonumber
e^{2A(r_{h})}=
\frac1{2\sqrt{g_{0}( g_1)^{3}}}\left(
r_h+ \frac32 \sqrt{ 1+4(p^1 g_1)/3 }\right)^{1/2}
\left(3r_h-\frac32 \sqrt{ 1+4(p^1 g_1)/3 }\right)^{3/2}=
\end{equation}
\begin{equation}
= \frac1{8 \sqrt{g_{0}( g_1)^{3}}}\left(
 \sqrt{1+4g_1 p^1} + 3 \sqrt{ 1+4(p^1 g_1)/3 }\right)^{1/2}
\left( 3\sqrt{1+4g_1 p^1}-3 \sqrt{ 1+4(p^1 g_1)/3 }
\right)^{3/2}\,.
\end{equation}
The solution represent a genuine black hole for a suitable choice of parameters $g_1$ and $p^1$. This choice corresponds to the requirement that the horizon shields the two singularities: these last are located at the points $r_{s,1} = -\frac32 \sqrt{ 1+4(p^1 g_1)/3 } $ and $r_{s,2} =\frac12 \sqrt{ 1+4(p^1 g_1)/3 }$, namely the zeros of the function $e^{2A(r)}$. Extensive details of the solution and the range of existence of a genuine black hole can be found in \cite{kiril_stefan}.

\subsection{\label{nonBPSsquareroot}Extremal non-BPS $F=-2i \sqrt{X^0(X^1)^3}$ solution}

We have seen  that the BPS states can be modified to become non-BPS ones by means of the clever trick described in Sec. \ref{rotation_charges}. In particular, an easy way to obtain a extremal non-BPS configuration is flipping the sign of one charge with respect to the BPS case. This corresponds to performing the trick of Sec. 2.2 using a  matrix $S=\pm\left(\begin{array}{cc} \textbf{1} &\ \ 0\\0&-\textbf{1}\end{array}\right)$. We start from ansatz for the metric and the form of the sections, that are the same as in the BPS case:
\begin{equation}
e^{2U(r)}= e^{\mathcal{K}}\frac{(r^2-r_h^2)^2}{r^2}\,, \qquad
e^{A(r)}= r e^{-\mathcal{K}/2}\,.
\end{equation}
\begin{equation}
 X_0 = \alpha_0 + \frac{\beta_0}{r}\,, \qquad X_1 = \alpha_1 + \frac{\beta_1}{r} \qquad z= \frac{X_1}{X_0}\,,
\end{equation}
\begin{equation}\label{using2}
\alpha_0= - \frac{1}{4 g_0}\,, \qquad \beta_0= -\frac{\xi_1 \beta_1}{\xi_0}\,, \qquad \alpha_1= - \frac{3}{4 g_1}\,, \qquad
r_h^2=\frac{16}{3}(g_1 \beta_1)^2-\frac12\,.
\end{equation}
At this point we perform the trick of flipping the sign of one charge, in particular we focus on the case in which the sign of $p^1$ is flipped, corresponding to a vector $\tilde Q$ given by
\begin{eqnarray}
\tilde Q=(\tilde p^0\ ,\tilde p^1)=(p^0\ ,-p^1) \ .
\end{eqnarray}
The relation \eqref{quant_ours}, valid in the BPS case, turns into this condition:
\begin{equation}
g_0p^0-g_1 p^1 =-1\,.
\end{equation}
Furthermore, the charges are written in terms of the other parameters as
\begin{equation}\label{pgb}
p^0=  - \frac{2}{g_0} \left(\frac18 +\frac{8 (g_1 \beta_1)^2}{3} \right)\,,\qquad
p^1= + \frac{2}{g_1} \left(\frac38 -\frac{8 (g_1 \beta_1)^2}{3} \right)\,.
\end{equation}
Alternatively, we can write the other parameters in terms of $p^1$
\begin{equation}
p^0=\frac{g_1 p^1 -1}{g_0}\,, \qquad \beta=-\frac{3 \sqrt{ 1-4(p^1 g_1)/3 }}{8 g_1}\,, \qquad r_h= \frac{\sqrt{1-4g_1 p^1} }{2}\,.
\end{equation}
The warp factor in this case turns out to be
\begin{eqnarray}\label{}
 e^{2U}&=&
\frac{2\sqrt{g_{0}( g_1)^{3}}(r^{2}-r_{h}^{2})^{2}}
{\left(r+ \frac{3}{2} \sqrt{ 1-4(p^1 g_1)/3 }\right)^{1/2}
\left(3r-\frac32 \sqrt{ 1-4(p^1 g_1)/3 }\right)^{3/2}} \,,
\nn\\ 
\end{eqnarray}
while the scalar profile is
\begin{equation}
z= z_{\infty} \sqrt{\frac{2r-\sqrt{ 1-4p^1 g_1/3 }}{2r+ 3 \sqrt{ 1-4p^1 g_1/3 }   }}\,.
\end{equation}
The non-BPS solution looks qualitatively similar to the BPS one for what concerns the general behaviour of the warp factor and the location of the singularities. The configuration depends on the parameters $g_0$, $g_1$ and $p^1$ (or alternatively, $\beta_1$), like in the BPS case. 

\subsubsection{Dirac quantization condition}

As mentioned before, the non-BPS solutions, in general, do not satisfy the Dirac quantization condition, but they do satisfy, by construction, another relation. Focusing on solution obtained by changing the sign of $p^1$ (described in the last section), the charges satisfy:
\begin{equation}
g_0 p^0-g_1 p^1=-1\,.
\end{equation}
As for the solution in section \ref{dir_nonBPS-tcube}, we have to impose, also in this case, a Dirac quantization condition with generic $n \in \mathbb{Z}$, from \eqref{dirac_ultima},
\begin{equation}\label{dir23}
g_0 p^0 +g_1 p^1 =n\,.
\end{equation}
This constraint, together with \eqref{pgb}, yields the relation 
\begin{equation}\label{rhmodel2}
g_{\Lambda} p^{\Lambda}= \frac12 -\frac{32}{3}(g_1 \beta_1)^2= -\frac{1}{2}- 2 r_h^2\,.
\end{equation}
Imposing the quantization condition \eqref{dir23} requires, also in this case,
\begin{equation}
-2 r_h^2 \in \mathbb{Z} + \frac12 \,.
\end{equation}  
To have proper black holes (finite nonzero area of the event horizon) the parameter $r_h$ has to be positive: this restricts the possible values of $n$ to be negative. Defining $m=-n$, we have then that only the values $m\in\bb N\backslash \{0,1\}$ correspond to proper black holes. The constraints of the magnetic charges are as follows:
\begin{equation}
g_0 p^0-g_1 p^1=-1\,,
\end{equation}
\begin{equation}
g_0 p^0+g_1 p^1=n =-m\,.
\end{equation}
This gives
\begin{equation}\label{quant_cond_cha}
2 g_0 p^0=-1-m \qquad 2 g_1 p^1=1-m\,.
\end{equation}
Note that the charges $p^0$ and $p^1$ are always negative; furthermore, notice that the configuration with $p^1=0$ is a naked singularity.
The scalar field solution is of the form:
\begin{equation}
z= z_{\infty} \sqrt{\frac{2r-\sqrt{3(2m+1)}}{2r +3 \sqrt{3(2m+1)}}}\,,
\end{equation}
and the warp factor reduces to:
\begin{equation}
 e^{2U}=
\frac{2\sqrt{g_{0}( g_1)^{3}}(r^{2}-r_{h}^{2})^{2}}
{\left(r+ \frac{3}{2} \sqrt{(2m+1)/3 }\right)^{1/2}
\left(3r-\frac32 \sqrt{(2m+1)/3 }\right)^{3/2}} \,.
\end{equation}
From \eqref{rhmodel2}  $r_h=\frac{\sqrt{2m-1} }{2}$ , and consequently the entropy takes the form 
\begin{equation}
e^{2A(r_h)}=2\sqrt{p^0 (p^1)^3}\, \frac{\left(r_h + \frac32 \sqrt{(2m+1)/3}\right)^{1/2}  \left(3 r_h - \frac32 \sqrt{(2m+1)/3}\right)^{3/2}}{\sqrt{(m^2-1)(1-m)^2} }=
\end{equation}
\begin{equation}
=\frac{3 \sqrt3}{2} \sqrt{p^0 (p^1)^3} \frac{ \left( \sqrt{2m-1} + \sqrt{ 3(2m+1)} \right)^{1/2}  \left( \sqrt{2m-1} - \sqrt{(2m+1)/3}\right)^{3/2} }{ |m-1| \sqrt{(m^2-1)} }\,.
\end{equation}
Notice that once again the prefactor $\sqrt{p^0 (p^1)^3}$ is the same one we find in the area formula of the corresponding ungauged supergravity configuration, with the same magnetic charges. This suggests an underlying duality structure of extremal solutions also in gauged Supergravities.

\subsection{\label{nonextr_squareroot}Nonextremal generalization of the magnetic configuration}

In this section we briefly recap the main features of the nonextremal generalization of the magnetic configurations previously described. Extensive details are provided in \cite{ours_nonextr}; this general nonextremal solution appeared also in \cite{klemm}.
The nonextremal deformation is characterized by:
\begin{equation}\label{nonextr_new_sol2}
e^{2U(r)}= e^{\mathcal{K}}\left( r^2+c - \frac{\mu}{r}+\frac{Q}{r^2} \right)\,, \qquad e^{A(r)}=e^{-\mathcal{K}/2}r\,.
\end{equation}
The form of the sections is unaltered with respect to the BPS case:
\begin{equation}
 X^0 = \alpha_0 + \frac{\beta_0}{r}\,, \qquad X^1 = \alpha_1 + \frac{\beta}{r} \qquad z= \frac{X^1}{X^0}\,,
\end{equation}
\begin{equation}\label{using}
\alpha_0= \pm \frac{1}{4 g_0}\,, \qquad \beta_0= -\frac{g_1 \beta}{g_0}\,, \qquad \alpha_1= \pm \frac{3}{4 g_1}\,,
\end{equation}
The other parameters for the warp factors are 
\begin{equation}
c=1-\frac{32}{3}(g_1 \beta)^2\,,
\end{equation}
\begin{equation}\label{mu2}
\mu= \frac83 \beta g_1 +\frac{512}{27} \beta^3 g_1^3 -3 \frac{ g_0^2 (p^0)^2}{4 \beta g_1}+ \frac{g_1 (p^1)^2}{12 \beta}\,,
\end{equation}
\begin{equation}\label{Q2}
Q=  g_0^2 (p^0)^2 + \frac13 g_1^2 (p^1)^2- \frac{16}{3} \beta^2 g_1^2 - \frac{256}{27} \beta^4 g_1^4 \,.
\end{equation}
We verified that there exist suitable sets of parameters such that the solution found represents a genuine nonextremal black hole. The singularities are located at $r_{s,1}= \pm 4 g_1 \beta $ and $r_{s,2}= \mp 4 g_1 \beta/3  $, and represent the zeros of the function $e^{-\mathcal{K}}$.
Also in this case, the physical configurations are those satisfying the Dirac quantization condition \eqref{dir23}.

\subsection{Product of the areas}

 The product of the areas for the nonextremal solutions above was already given in \cite{ours_nonextr}. As mentioned before, once again we take the product over the four roots of the warp factor. The relevant quantities $\kappa_1$ and $\kappa_2$ are:
\begin{equation}
\kappa_1 = (2 g_0 p^0)^2 \qquad \kappa_2= (2 g_1 p^1)^2\,.
\end{equation}
Finally the product of the four areas results in
\begin{equation}\label{prod2}
 \prod_{\alpha=1}^4 A_{\alpha}= (4 \pi)^4 27 \frac{p^0 (p^1)^3}{g_0 (g_1)^3}\,.
\end{equation}
We have still to impose the Dirac quantization condition \eqref{quant_cond_cha}. If we do so, we can express the product of the areas as
\begin{equation}
 \prod_{\alpha=1}^4 A_{\alpha}= (4 \pi)^4 \frac{27}{16} \frac{(m+1)^2(m^2-1)}{g_0^2 (g_1)^6}\,.
\end{equation}

\section{Mass of the black hole solutions}

We are now going to compute the mass of the various black hole solutions found in the previous sections\footnote{Throughout the section we redefine $\beta_1\equiv\beta$, also for the extremal case.}. In Appendix \ref{mass-formulae} is explained that the formalism developed in \cite{ours_BPSbound} provides the mass (as quantity appearing in the superalgebra) for configurations that satisfy \eqref{quant}, such that the state asymptotically preserves some supersymmetry. When indeed \eqref{quant} is satisfied, the mass of the dyonic nonextremal solution of Section \ref{3} turns out to be:
\begin{equation}\label{result_mass2}
M=-\frac{ (-3 + 2 g^1 (24 \beta^2 g^1 \mp q_1)) (-3 + 4 g^1 (48 \beta^2 g^1 \mp q_1))}{72 \sqrt2 \beta ( g_0 (g^1)^7)^{1/4}}\,.
\end{equation}
For the magnetic nonextremal configurations of Section \ref{4}, instead, the mass is
\begin{equation}\label{result_mass3}
M =- \frac{ \big( -9 +2 g_1 (8 \beta^2 g_1 \pm 3 p^1) \big)  \big( -9+4g_1 (16 \beta^2 g_1 \pm 3p^1) \big) }{ 72  \sqrt{2} \,3^{1/4}  \beta (g_0 g_1^7)^{1/4}  } \,,
\end{equation}
and was already computed in \cite{ours_nonextr}.
For all the nonextremal black hole solutions we found so far the mass computed with \eqref{result_mass2} and  \eqref{result_mass3} turns out to be positive. Furthermore, the mass is zero if computed on the BPS configuration, as it should be, since a supersymmetric state saturates the BPS bound $M\geq0$  \cite{ours_BPSbound} \cite{kiril}. 

The mass computation for the configurations that do not satisfy
\begin{equation}\label{1}
g_{\Lambda}p^{\Lambda}-g^{\Lambda}q_{\Lambda}= \pm 1 \,
\end{equation}
represent instead a more challenging issue. In this case the mass formula \eqref{mass} gives a divergent result. For instance, the non-BPS solutions of sections  \ref{non-BPS-t3-extremal} and \ref{nonBPSsquareroot}  fall in this category.

For these states we can try and use the other formula given in \cite{kiril}, namely \eqref{mass_ele}. This formula gives the mass appearing in the superalgebra when we are dealing with solutions that asymptote to ordinary AdS$_4$ spacetime, and in that case it coincides with the one obtained via holographic renormalization \cite{batra}. After suitable rescaling (details can be found  in \cite{kiril}) it gives for the nonextremal configuration
\begin{equation}
M_{holo}= \frac{288 \beta^2 g_1^2 -1024 \beta^4 g_1^4-81 g_0^2 (p^0)^2 +9 g_1^2 (p^1)^2}{108 \sqrt2 3^{3/4} \beta g_1 g^2}(g_0 g_1^3)^{1/4}\,.
\end{equation}
In the extremal BPS case this boils down to:
\begin{equation}
M_{holo}^{BPS}= -\frac{128 \sqrt2  \beta^3 g_1^3 (g_0 g_1^3)^{1/4}}{9 \times 3^{3/4}g^2}\,.
\end{equation}
The result has the same dependence as the one found in \cite{kiril_stefan} (the normalization is different). The mass formula for the dyonic solution can be inferred from this one by performing the symplectic rotation.
Being quadratic in the charges, the holographic mass formula gives the same result for the non BPS extremal solution and for the BPS one. This formula, though, does not reproduce the BPS bound found in \cite{ours_BPSbound}, since for the BPS configurations gives a value that in general is nonzero.

Another comment here is in order: the divergent part of the mass formula \eqref{mass} result is proportional to $g_{\Lambda} p^{\Lambda} - g^{\Lambda}q_{\Lambda} \pm 1$. If we subtract the mass result of two configurations with the same divergence we get in principle a finite quantity. We can then introduce a quantity $M_{gap}$ that turns out to be finite, given by
\begin{equation}\label{22}
{M_{gap}}^{NON\,\, BPS} ={M_{non-extr}}^{NON \,\, BPS}-{M_{extr}}^{NON\,\,BPS} \,.
\end{equation}
In this case the difference is taken between the nonextremal deformation over the non-BPS state and the non-BPS extremal black hole. One should be careful in interpreting such quantity, though: it is finite but it is extracted from quantities that are per se divergent. 
If we proceed with the computation, we find, for the nonextremal dyonic black holes satisfying  $g_0 p^0+g^1 q_1=-1$ (we have already imposed the quantization condition \eqref{dirac_ultima})
\begin{equation}
{M_{gap}}^{NON\,\, BPS}=(p^0 q_1^3)^{1/4} \frac{  -2304 \beta^4 (1 + m)^4 + 
 144 \beta^2 m (1 + m)^2 q_1^2 + (1 + m (-7 + 10 m)) q_1^4}{72 \sqrt2  ((1-m) (1+m)^{7})^{1/4} \beta q_1^3}\,.
\end{equation}
The result for the magnetic configurations satisfying $g_0 p^0-g_1 p^1=-1$ is:
\begin{equation}\label{guess}
{M_{gap}}^{NON\,\, BPS} =(p^0 (p^1)^3)^{1/4} \frac{-256 \beta^4 (1 + m)^4 + 144 \beta^2 m (1 + m)^2 (p^1)^2 + 
 9 (1 - 7 m + 10 m^2) (p^1)^4}{ 72 \sqrt2 \, 3^{1/4} ((m-1) (1+m)^7)^{1/4} \beta (p^1)^3} \,.
\end{equation}

To sum up, both the options for computing the mass/mass gap of the not asymptotically-BPS black holes present some issues, in different ways. For these non-BPS configurations the definition of the mass requires a more careful and deep examination. It would be interesting to have a consistent framework for defining such quantities that does not rely on the asymptotic supersymmetry of the solution, but at the same time reproduces the correct BPS bound.

\section*{Acknowledgements}

We would like to thank Gianguido Dall'Agata and Stefan Vandoren for encouraging our work on this project, and for precious discussions that lead important insights on the results presented in this paper. We also thank Kiril Hristov for useful comments and discussions.
We acknowledge support by the Netherlands Organization for Scientific Research (NWO) under the VICI grant 680-47-603.

\appendix

\section{\label{2ndBPS-branch}General choice for the Killing spinor projectors}

\addtocontents{toc}{\setcounter{tocdepth}{-1}}
\subsection{First set of BPS equations}

In order to solve the equations of the gravitino and gaugino field, the SUSY parameters get constrained by two projection conditions. They have been written in \cite{Dall'Agata:2010gj} as
\begin{eqnarray}\label{projs}
\g^0\e_A&=&i e^{i\alpha}\varepsilon_{AB}\e^B\ ,\nn\\
\g^1\e_A&=&e^{i\alpha}\delta_{AB}\e^B\ ,
\end{eqnarray}
and give rise to the set of Supersymmetry first order equations studied there.
This is not the most general choice of constraints, since a different phase between the two equations is also possible. We revise here what the allowed choices are, and how they are determined. They correspond to two sets of Killing spinors and BPS equations, that were presented in  \cite{kiril_stefan} for a particular choice of $\gamma$-matrices convention. 

\subsection{Choice of the phase}

We work in the signature $(-,+,+,+)$
and use the conventions
\begin{eqnarray}
\e_{0123}&=&1 \qquad \gamma_5=-i\g_0\g_1\g_2\g_3 \qquad \g_5\e_A=\e_A \qquad \g^5\e^A=-\e^A\ ,
\end{eqnarray}
where $\epsilon_A$ and $\epsilon^A$ are \textcolor{red}{} two Weyl spinors of opposite chirality .
We take $\e_{abcd}\g^{cd}=2i\g_{ab}\g^5$, $\g^5=i\g^0\g^1\g^2\g^3$. The identity matrix is built as
\begin{eqnarray}
\delta_{AB}&=&(i\sigma^2)_A^C\,\e_{BC}
\end{eqnarray}
and satisfies $\delta^{AB}=\delta_{AB}$, while the antisymmetric tensor is $\epsilon_{AB}=-\epsilon_{BA}=\epsilon^{AB}$ . \\
We define two projectors as
\begin{eqnarray}\label{projs}
\g^0\e_A&=&i e^{i \eta}\varepsilon_{AB}\e^B\ ,\nn\\
\g^1\e_A&=&e^{i \mu}\delta_{AB}\e^B\ ;
\end{eqnarray}
compatibility of these two definitions requires that we also take
\begin{eqnarray}
\g^0\e^A&=&-i\,e^{-i \eta}\varepsilon^{AB}\e_B\ ,\qquad \g^1\e^A=e^{-i \mu}\delta^{AB}\e_B\ .
\end{eqnarray}

Moreover, the definition above has to be consistent with the properties of the two $\gamma$-matrices. In particular, consider the anticommutation  $\{\gamma^0,\gamma^1\}=0$. By looking at the action on the to-be-Killing spinors, namely $\g^0\g^1\e_A$ and $\g^1\g^0\e_A$ , we have that
\begin{eqnarray}
\g^0\g^1\epsilon_A&=&-ie^{i(\mu-\eta)}\delta_{AB}\varepsilon^{BC}\epsilon_C\ ,\nn\\
\g^1\g^0\epsilon_A&=&i e^{i(\eta-\mu)}\varepsilon_{AB}\delta^{BC}\epsilon_C\ ,
\end{eqnarray}
which implies, for consistency, that
\begin{eqnarray}
\eta-\mu=k\pi\ ,\qquad k\in\bb Z\ .
\end{eqnarray}
This means that we can restrict the form of the projectors in \eqref{projs} to the choice of a single phase $\alpha$, by setting $\eta\equiv \alpha + k \pi$, $\mu\equiv \alpha$, and precisely
\begin{eqnarray}\label{restr-projs_star}
\g^0\e_A&=&\pm i e^{i \alpha}\varepsilon_{AB}\e^B\ ,\nn\\
\g^1\e_A&=& e^{i \alpha}\delta_{AB}\e^B\ .
\end{eqnarray}
The two choices of $\pm1$, in the equations above, give two BPS branches of SUSY equations. 

\subsection{Second BPS branch}

The \lq{}\lq{}$+$"  case was analyzed in \cite{Dall'Agata:2010gj}, we will study the supersymmetry equation for the other case in the following.

\textit{We show that the \lq{}\lq{}$-$\rq{}\rq choice can be absorbed in a \lq\lq{}sign\rq\rq{} redefinition of black hole charges, which thus correspond to $\mc Z\rar-\mc Z$.}

Consider the projector conditions
\begin{eqnarray}\label{restr-projs}
\g^0\e_A&=&-i e^{i\alpha}\varepsilon_{AB}\e^B\ ,\nn\\
\g^1\e_A&=& e^{i\alpha}\delta_{AB}\e^B\ ,
\end{eqnarray}
and the corresponding relations
\begin{eqnarray}
\g^0\e^A&=&i\,e^{-i\alpha}\varepsilon^{AB}\e_B\ ,\qquad \g^1\e^A=e^{-i \alpha}\delta^{AB}\e_B\ .
\end{eqnarray} 

We seek black holes solutions in a zero fermions background, thus we require $\delta \psi^{A}_{\mu}=0$ and $\delta \lambda^{iA} = 0$.
The SUSY variations for the gravitino and gaugino, in $\mc N=2$ $U(1)$-gauged SUGRA are \cite{Andrianopoli:1996cm}
\begin{eqnarray}
	\label{trsfSUSY}
\delta\psi_{\mu\,A}&=&D_{\mu}\epsilon_{A}-\varepsilon_{AB}\, T^{-}_{\mu\nu}\,\gamma^{\nu}\,\epsilon^{B}+\frac{i}{2}\,{\cal L}\, \delta_{AB}\,\gamma^{\nu}\,\eta_{\mu\nu}\,\epsilon^{B}\ ,
\\[2mm]
\delta\lambda^{iA}&=&-i\,\partial_{\mu}z^{i}\,\gamma^{\mu}\,\epsilon^{A}+G^{-i}_{\mu\nu}\,\gamma^{\mu\nu}\,\varepsilon^{AB}\,\epsilon_{B}+\overline{D}^{i}\overline{\cal L}\,\delta^{AB}\,\epsilon_{B}\ ,
\end{eqnarray}
where the covariant derivative is
\begin{eqnarray}
	D_{\mu}\epsilon_{A}\equiv\partial_{\mu}\epsilon_{A}-\frac14\,\omega_\mu^{ab} \gamma_{ab}\epsilon_{A}+\frac{i}{2}\,{\cal A}_{\mu}\epsilon_{A} +\langle \mc G\,, \mathbb A_\mu
\rangle\, \delta_{AC}\varepsilon^{CB}\epsilon_B\ ,
\end{eqnarray}
with $\mc A_{\mu}$ being the K\"ahler transformations connection
\begin{equation}
	{\cal A}_\mu \equiv \frac{i}{2}\, \left(\partial_\mu\bar z^{\bar \jmath}\, \overline{\partial}_{\bar\jmath}K - \partial_\mu z^i\, \partial_i K\right)\ .
\end{equation}
In order to write a covariant expression for the supersymmetry transformations, we introduced the vector 
\begin{eqnarray}
\mathbb{A}_{\mu}=\left(
\begin{array}{c}
A^{\Lambda}_{\mu}\\
 \tilde A_{\mu\, \Lambda}
\end{array}
\right)
\end{eqnarray}
whose components are the electromagnetic potentials and their dual ones \cite{deWit:2005ub}. The ansatz for the field strengths $F_{\mu\nu}^\Lambda = 2\partial_{[\mu}A_{\nu]}^\Lambda$ is
\begin{eqnarray}\label{vectoransatz}
F^{\Lambda}_{tr}&=&\frac{e^{2U-2\psi}}{2} \, ({\cal I}^{-1})^{\Lambda\Sigma}\, \left(  {\cal R}_{\Sigma \Gamma}\,p^{\Gamma}-q_{\Sigma} \right)\ ,\\[4mm]
F^{\Lambda}_{\theta\phi}&=&-\frac12p^{\Lambda}\sin\theta\ .
\end{eqnarray}
They appear in the SUSY variations both via their (anti)self--dual combinations
\begin{equation}
	F_{\mu\nu}^- \equiv \frac12 \left(F_{\mu\nu} - \frac{i}{2}\epsilon_{\mu\nu\rho\sigma} F^{\rho \sigma}\right),
\end{equation}
and dressed by the scalar fields
\begin{equation}
\label{trsfPieces}
T^{-}_{\mu\nu}=2i \, {\cal I}_{\Lambda\Sigma}\, L^{\Sigma}\, {F}^{\Lambda-}_{\mu\nu}\, \qquad \qquad
G^{-i}_{\mu\nu}=\overline{D}^i\bar L^{\Gamma}\, {\cal I}_{\Gamma\Lambda}\,F^{\Lambda-}_{\mu\nu}.
\end{equation}

\begin{itemize}
\item\noindent
The variation of the gravitino time-component gives the first order equation  $\delta \psi_{t A} = 0$, which, for a time independent Killing spinor, corresponds to 
\begin{equation}
	-\frac12 e^{2U}U'\gamma^{01}\epsilon_{A}+ \frac12\, A_t^\Lambda g_{\Lambda} \delta_{AC} \varepsilon^{CB} \epsilon_B +\frac {i}{2}\,e^{3U-2\psi}\,{\cal Z}\,\gamma^{1}\varepsilon_{AB}\epsilon^{B}-\frac{i}{2}\,e^U\,{\cal L}\,\delta_{AB}\gamma^{0}\epsilon^{B}=0\ .
\end{equation}
Using the projection conditions \eqref{restr-projs} this equation reduces to
\begin{eqnarray}
(-U\rq{}-ie^{-2U}A_t^\Lambda g_{\Lambda}+e^{U-2\psi}e^{-i\alpha}\mc Z-ie^{-U}e^{-i \alpha}\mc L)\e_A&=&0\ ,
\end{eqnarray}
whose real and imaginary part give the first order flow equation for the warp factor $U$ and the constraint on the gauge fields:
\begin{eqnarray}\label{eqnU}
U\rq{}&=&e^{U-2\psi}\re(e^{-i\alpha}\mc Z)+e^{-U}\im(e^{-i \alpha}\mc L)\ ,\nn\\
-e^{-2U}\, \langle \mc G\,, \mathbb A_t
\rangle&=&-e^{U-2 \psi} {\rm Im}(e^{-i \alpha}{\cal Z})+e^{-U}{\rm Re}(e^{-i \alpha}{\cal L}).
\end{eqnarray}
\item
The radial component $\delta\psi_r=0$ yields
\begin{equation}
\partial_{r}\epsilon_{A}+\frac i2  {\cal A}_r\epsilon_{A} -\frac i{2R^{2}}e^{U-2\psi} {\cal Z}\gamma^{0}\varepsilon_{AB}\epsilon^{B}+\frac i2{\cal  L}\,\delta_{AB}\gamma^{1}e^{-U}\epsilon^{B}=0\ ,
\end{equation}
which reduces to
\begin{equation}
	\partial_{r}\epsilon_{A}-\frac12 \, \left(U' -  i \widetilde {\cal A}\right)\epsilon_{A} =0,
\end{equation}
where we introduced
\begin{equation}
	\widetilde{\cal A} = {\cal A}_r -\left(e^{U-2 \psi}\,{\rm Im}(e^{-i \alpha} {\cal Z})-e^{-U}\, {\rm Re}(e^{-i \alpha}\mc L)\right).
\end{equation}
This equation is readily solved by 
\begin{equation}\label{epsradial}
	\epsilon_A = e^{\frac{U}{2} - \frac{i}{2}\int \widetilde{\cal A} \,dr}\chi_A,
\end{equation}
for a spinor $\chi_A$ that is $r$ independent and satisfies\ \ 
$
	\gamma^0 \chi_A =-i\, \varepsilon_{AB} \chi^B, \quad \gamma^1 \chi_A = \delta_{AB} \chi^B.
$
Let us apply any of the projection conditions defined above to the Killing spinors whose radial dependence have been fixed as in \eqref{epsradial}. We have, for instance, that
\begin{eqnarray}
\g^0\chi_A&=&ie^{i\left(
\alpha+\int \widetilde{\cal A} \, dr
\right)}\varepsilon_{AB}\chi^B\ .
\end{eqnarray}
Since $\chi_A$ and $\chi^A$ are constant in the r-coordinate, then
\begin{eqnarray}
	\alpha +\int \widetilde{\cal A} \, dr &=& const\ ,
\end{eqnarray}
from which we derive the flow equation for the phase
\begin{equation}
	\alpha'+{\cal A}_r = e^{U-2 \psi} \, {\rm Im} (e^{-i \alpha}{\cal Z}) - e^{-U}\, {\rm Re}(e^{-i \alpha}{\cal L}) \ .
\end{equation}
\item The angular part of the gravitino variation in the $\theta$  direction gives
\begin{equation}
	\partial_{\theta}\epsilon_{A}+\frac12\, e^{\psi}(U'-\psi')\gamma^{12}\epsilon_{A}-\frac12\,e^{U-\psi}\,{\cal Z}\,
	\gamma^{3}\varepsilon_{AB}\epsilon^{B}+\frac i2\,e^{-U+\psi}\,
	{\cal L}\,\delta_{AB}\gamma^{2}\epsilon^{B}=0.
\end{equation}
This is a little bit trickier, but we can easily see that
\begin{eqnarray}
\varepsilon_{AB}\e^B&=&ie^{-i\alpha}\g^0\e_A\ ,\nn\\
\delta_{AB}\e^B&=&e^{-i \alpha}\g^1\e_A\ ,\nn\\
\g^3\g^0&=&i\g^2\g^1\g^5\ ,
\end{eqnarray}
and the equation can be written as
\begin{eqnarray}
\partial_{\theta}\epsilon_{A}+\frac12\, e^{\psi}\left[
U\rq{}-\psi\rq{}-e^{U-2\psi}(e^{-i\alpha}\mc Z)-ie^{-U}(e^{-i \alpha}\mc L)\,
\right]\g^{12}\e_A&=&0\ ;
\end{eqnarray}
we can also use the $U$ flow equation, \eqref{eqnU}, so this boils down to
\begin{equation}
	\partial_{\theta}\epsilon_{A} = \frac12\, e^\psi\,\left[\psi'-2 e^U\,{\rm Im}(e^{-i \alpha}{\cal L})+i\left(e^{U-2 \psi}\,{\rm Im}(e^{-i \alpha} {\cal Z})+e^{-U}\,{\rm Re}(e^{-i \alpha}{\cal L})\right)\right]\gamma^{12} \epsilon_A.
\end{equation}
The radial dependence of the Killing spinor has been determined earlier by (\ref{epsradial}). This means that the quantity between square brackets, which has a radial dependence, is required to vanish. This yields the flow equation for $\psi$ 
\begin{equation}
	\psi'=2  e^U\,{\rm Im}(e^{-i \alpha}{\cal L})\ ,
\end{equation}
and the constraint
\begin{equation}
	e^{U-2 \psi}\,{\rm Im}(e^{-i \alpha} {\cal Z})=-e^{-U}\,{\rm Re}(e^{-i \alpha}{\cal L})\ ,
\end{equation}
which defines the phase $\alpha$ as
\begin{equation}
  e^{2 i \alpha} = \frac{{\cal Z} + i\, e^{2(\psi-U)}{\cal L}}{\overline{\cal Z} - i\, e^{2(\psi-U)}\overline{\cal L}}\ .
\end{equation}
This also fixes the ansatz for the time component of the vector fields
\begin{equation}\label{def-alpha}
	\langle\mc G\,,\mathbb{A}_t\rangle = -2\, e^U\, {\rm Re}(e^{-i \alpha}{\cal L}).
\end{equation}
We finally get that the Killing spinors $\epsilon_A$ should not depend on $\theta$:
\begin{equation}
	\partial_{\theta} \epsilon_A = 0.
\end{equation}
\item
A further constraint comes from the angular component of the gravitino variation in the $\theta$ direction.
With the redefined projector \eqref{restr-projs}, the equation gets an extra \lq\lq{}$-$\rq\rq{} sign in front of the symplectic product of black hole and gauge charges, with respect to the case analyzed in \cite{Dall'Agata:2010gj},  giving precisely
\begin{eqnarray}
	\partial_{\phi}\epsilon_A &=&\frac12\,\cos \theta\, \gamma^{23} \epsilon_A + \frac i2\, \langle {\cal G}, Q\rangle \, \cos \theta\, \gamma^{01} \epsilon_A\ ,
\end{eqnarray}
and then a charge quantization condition of the form
\begin{eqnarray}\label{QCond-BPS2}
\langle {\cal G}, Q\rangle =1\ .
\end{eqnarray}
\item Given the constraints and equations obtained so far, the dilatino variation $\delta \lambda^{iA} = 0$ eventually gives the flow equations for the scalar fields in the form
\begin{equation}
	z^i{}' =  e^{i \alpha} g^{i \bar \jmath}\left[e^{U -2 \psi}\, \overline D_{\bar \jmath} \overline{\cal Z}  - i \, e^{-U}\, \overline D_{\bar \jmath} \overline{\cal L}\right]\ .
\end{equation}

\end{itemize}

By comparing this analysis of BPS equations with the one in \cite{Dall'Agata:2010gj}, one can easily see that a black hole solution can be constructed by taking a 1/4-BPS one, and flipping the sign of the electric and magnetic charges. The new configuration does not satisfy the BPS equation derived from choice of \lq\lq{}$+$\rq\rq{} in \eqref{restr-projs_star}, but instead it satisfies the BPS equations wrt the choice of the \lq\lq{}$-$\rq\rq{} sign, so it is still a BPS solution of the $\mc N=2$ U(1)-gauged theory.

\subsection{Comment on the independent BPS branches}
As we recalled earlier in the paper, flipping the sign of the black hole charges for a BPS solution in the ungauged theory would produce another \textit{equivalent} supersymmetric solution, satisfying the same ungauged flow equations. This is due to the fact that the phase of the projector of the Killing spinor is defined up to a phase $\pi$, both in the ungauged and the guged  case  (see, for the latter case, eq. \eqref{def-alpha}).

In absence of gauging, if we flip the sign of $Q$, and thus of $\mc Z$, the change of sign in the flow equations can be absorbed in the shift of $\alpha\rar\alpha\pm\pi$. 
When $\mc G\neq0$, the shift of the phase absorbs  the  flip of sign of both $Q\rar-Q$ and $\mc G\rar-\mc G$  \textit{simultaneously}, so that the branches possibly generated by this operation are already taken into account, once we study configurations with unrestricted charges. 

However, if one only changes the sign of one vector of charges, $Q$ \textit{or} $\mc G$, a second set of BPS equation is produced, which is physically inequivalent to the previous one, given that the physical black hole and gravitino charges for the second branch have opposite orientation. In fact, the black hole charges are constrained, together with the gravitino charges, to satisfy the quantization condition \eqref{QCond-BPS2}, and one sees that the sign of the rhs changes for the two branches.

In conclusion, in order not to miss any supersymmetric configuration of the theory, one has to consider both sets of BPS equations, namely those generated by the \eqref{projs} and \eqref{restr-projs}, each one with the corresponding quantization condition,
\begin{eqnarray}
\langle\mc G\ , Q\rangle_{1^{st}BPS} =-1\ ,\qquad \langle\mc G\ , Q\rangle_{2^{nd}BPS}  =1\ ,
\end{eqnarray}
and the relative physical assignment of black hole and gravitino charges.

\addtocontents{toc}{\setcounter{tocdepth}{2}}

\section{\label{mass-formulae}Mass formulas}

In this section we would like to collect and review the formulas at our disposal to compute the mass for the black hole configurations found so far. 

These mass formulas were found first in \cite{ours_BPSbound} and \cite{kiril}, where conserved quantities like the mass of the configuration were read off directly from the superalgebra. We can generalize these expressions found to mass formulas valid in case of both $g_{\Lambda}$ and $g^{\Lambda}$ terms, due to the symplectic invariance of the supersymmetry variations.  
The generalization consist in:
$$
M= \frac{1}{8 \pi} \lim_{r \rightarrow \infty} \oint {\rm d}\Sigma_{tr} \left( g' r+ \frac{g'}{2 g^2 r}\right)  \bigg( 2 {\text Im}(L^{\Lambda} q_{\Lambda} -M_{\Lambda}p^{\Lambda}) \sin \theta e_0^t e_1^r e_2^{\theta} e_3^{\varphi} +
$$
\begin{equation}
+2g | P_{\Lambda}^a L^{\Lambda}- P^{\Lambda |a} M_{\Lambda}|e_0^t e_1^{r}  - (\omega_{\theta}^{12} e_0^t e_1^{r} e_2^{\theta} +\omega_{\varphi}^{13} e_0^t e_1^{r} e_3^{\varphi}) \bigg)\,. 
\end{equation}
This mass formula is valid for configurations whose charges satisfy
\begin{equation}\label{qu}
 g_{\Lambda} p^{\Lambda} - g^{\Lambda} q_{\Lambda} =-1\,.
\end{equation}
For the BPS  (dyonic or magnetic) extremal black holes presented in sections \ref{t3BPS} and \ref{squareroot_BPS} the mass turns out to be zero:
\begin{equation}
M=0\,.
\end{equation}
This is a consequence of that fact that supersymmetric configuration saturate the bound $M \geq 0$ \cite{ours_BPSbound}. However, for a generic non-supersymmetric configuration, like the non-BPS ones (sections \ref{non-BPS-t3-extremal} and \ref{nonBPSsquareroot}) or the non-extremal ones (sections \ref{nonextr_tcube} and \ref{nonextr_squareroot}), the mass formula in general does not give a finite result,  the divergent part being proportional to $(g_{\Lambda} p^{\Lambda} - g^{\Lambda} q_{\Lambda} +1)$. Just for the configurations with charges that satisfy $g_{\Lambda} p^{\Lambda} - g^{\Lambda} q_{\Lambda} =-1\,$ the mass has a finite value.

In addition to the BPS solutions satisfying \eqref{qu}, we also have another branch of BPS solutions, corresponding to flipping all the sign of the charges. These are the solutions described in App. A, denoted by "second branch". In this case the quantization condition is \eqref{QCond-BPS2}, namely
\begin{equation}\label{quant_2}
g_{\Lambda} p^{\Lambda} - g^{\Lambda} q_{\Lambda} =1\,.
\end{equation} 
For this solution the following mass formula should be used:
$$
M= \frac{1}{8 \pi} \lim_{r \rightarrow \infty} \oint {\rm d}\Sigma_{tr} \left( g' r+ \frac{g'}{2 g^2 r}\right)  \bigg( - 2 {\text Im}(L^{\Lambda} q_{\Lambda} -M_{\Lambda}p^{\Lambda}) \sin \theta e_0^t e_1^r e_2^{\theta} e_3^{\varphi} +
$$
\begin{equation}
+2g | P_{\Lambda}^a L^{\Lambda}- P^{\Lambda |a} M_{\Lambda}|e_0^t e_1^{r}  - (\omega_{\theta}^{12} e_0^t e_1^{r} e_2^{\theta} +\omega_{\varphi}^{13} e_0^t e_1^{r} e_3^{\varphi}) \bigg)\,. 
\end{equation}
The minus sign in the first term in the integral is due to the different projections satisfied by the Killing spinor of the solution taken into consideration. Once computed with this formula, the mass turns out to be zero for the BPS states. Again, just the configurations with charges satisfying \eqref{quant_2} have a finite mass.

To sum up, we are able to compute masses for black holes satisfying the Dirac quantization condition for integer numbers $\pm1$. The outcome of the mass formula in this case corresponds to a finite conserved quantity. The formula for the two signs respectively is:
$$
M= \frac{1}{8 \pi} \lim_{r \rightarrow \infty} \oint {\rm d}\Sigma_{tr} \left( g' r+ \frac{g'}{2 g^2 r}\right)  \bigg( \mp 2 {\text Im}(L^{\Lambda} q_{\Lambda} -M_{\Lambda}p^{\Lambda}) \sin \theta e_0^t e_1^r e_2^{\theta} e_3^{\varphi} +
$$
\begin{equation}\label{mass}
+2g | P_{\Lambda}^a L^{\Lambda}- P^{\Lambda |a} M_{\Lambda}|e_0^t e_1^{r}  - (\omega_{\theta}^{12} e_0^t e_1^{r} e_2^{\theta} +\omega_{\varphi}^{13} e_0^t e_1^{r} e_3^{\varphi}) \bigg)\,. 
\end{equation}

For completeness in this section we give also the mass formula for black holes that asymptote to ordinary AdS$_4$. It looks different from the previous one:
$$
M= \frac{1}{8 \pi} \lim_{r \rightarrow \infty} \oint {\rm d}\Sigma_{tr} \bigg(e_{[0}^t e_{1}^r e_{2]}^{\theta} + \sin \theta e_{[0}^t e_{1}^r e_{3]}^{\varphi} +2g g' r | P_{\Lambda}^a L^{\Lambda}- P^{a| \Lambda}M_{\Lambda}|e_{[0}^t e_{1]}^{r}+
$$
\begin{equation}\label{mass_ele}
  - \sqrt{g'^2 r^2 +1} (\omega_{\theta}^{12} e_{[0}^t e_1^{r} e_{2]}^{\theta} +\omega_{\varphi}^{13} e_{[0}^t e_1^{r} e_{3]}^{\varphi}) \bigg) \,.
\end{equation}
Further details about the superalgebras (and consequently, mass formulas) underlying solutions with different asymptotics can be found in \cite{ours_BPSbound} and \cite{kiril}.

\section{\label{EquivPrep}Equivalent prepotentials for the $t^3$ model and matching of the solutions}

The $t^3$-model, which parametrizes the coset $SU(1,1)/U(1)$, can be equivalently derived by two choices of the prepotential
\begin{itemize}
\item Cubic prepotential
\begin{eqnarray}\label{F1}
F(X)&=&\frac{(X^1)^3}{X^0}
\end{eqnarray}
\item Square-root prepotential
\begin{eqnarray}\label{F2}
F(X)&=&-2i\sqrt{X^0(X^1)^3}
\end{eqnarray}
\end{itemize}
The two parametrizations are equivalent, as we are going to show in detail in the rest of this section.

Let us first take $F$ as in \eqref{F1}, and identify the scalar field as
 $\frac{X^1}{X^0}\equiv -i\,t \ ;$
 we obtain the holomorphic sections 
\begin{eqnarray}
\mc V&=&\left(
\begin{array}{c}
1\\-i\,t\\-i\,t^3\\-3\,t^2
\end{array}
\right)\ .
\end{eqnarray}
The moduli space is defined by $\re t<0$ so, writing $t=\lambda+i\alpha$, this corresponds to the requirement $\lambda>0$. Axions are $\alpha=\im t$, in this choice of parameterization.

If we instead start from the superpotential \eqref{F2} and identify $z=\frac{X^1}{X^0}$,  the holomorphic sections are
\begin{eqnarray}
\tilde{\mc V}&=&\left(
\begin{array}{c}
1\\ z\\-i\,z^{3/2}\\-3i\,\sqrt{z}
\end{array}
\right)\ .
\end{eqnarray}

The two prepotentials being equivalent means that a real symplectic matrix $\mathrm B\in Sp(4,\bb R)$ exists, that rotates
\begin{eqnarray}
\tilde{\mc V}&=&\mathrm B \mc V\ ,\qquad \quad \tilde{\mc M}=\mathrm B^T\mc M\mathrm B\ ,
\end{eqnarray}
up to a holomorphic coordinate transformation.

It is easy to see that such matrix is
\begin{eqnarray}\label{ss}
\mathrm B&=&\left(
\begin{array}{cc|cc}
1&0&0&0\\
0&0&0&-1/3\\ \hline
0&0&1&0\\
0&3&0&0
\end{array}
\right)\ ,
\end{eqnarray}
which requires the identification 
\begin{eqnarray}\label{ident}
z\equiv(t)^2=(\lambda^2-\alpha^2)+2i\alpha\lambda\ .
\end{eqnarray}

If we now want to look at zero axions solutions, we have to set $\alpha\equiv0$, that is $\im \, t=0$, which corresponds to take $t=\lambda$. This is consistent with the correspondence above if we take  $z=\lambda^2$. Indeed, in the square-root prepotential model, the zero axion limit is obtained by taking $\im \, z=0$ in the scalar domain $\re \, z>0$, which in \eqref{ident} selects $z=\lambda^2$, as required.

\hspace{2mm}

In the light of this correspondence, one can easily see the mathcing between the solutions of sections 3 and 4. We show here the explicit rotation for the BPS case: the magnetic BPS solution described in section \ref{squareroot_BPS} can be rotated into the dyonic BPS solution of section \ref{t3BPS} by means of the (inverse of the) symplectic rotation \eqref{ss}.

Starting from the magnetic solution of the form \eqref{e2U} and \eqref{param_x}, whose parameters appearing in the sections are\footnote{This solution corresponds to the BPS solution of \cite{kiril_stefan} with $g_{\Lambda}p^{\Lambda}=-1$ and $\alpha = \pi$, where $\alpha$ is the phase of the Killing spinor. For this reason the signs of the parameters $\alpha^{\Lambda}$ are flipped with respect to the ones present in the literature \cite{kiril_stefan}.}
\begin{equation}\label{param_prima}
\alpha_0= \frac{1}{4 g_0}\,, \qquad \beta_0= -\frac{g_1 \beta}{g_0}\,, \qquad \alpha_1= \frac{3}{4 g_1}\,, \qquad
r_h= \frac{\sqrt{1+4g_1p^1}}{2}\,.
\end{equation}
we perform the inverse symplectic transformation $\mathrm{B}^{-1}$. The transformation acts on the gauging charges and on the electric/magnetic charges. The rotated quantities (denoted with $'$) are:
\begin{equation}
{g^0}'=g^0\,, \qquad {g^1}'=\frac13 g_1\,, \qquad g_0'=g_0\,, \qquad g_1'=-3 g^1\,,
\end{equation}
and
\begin{equation}
{p^0}'=p^0\,, \qquad {p^1}'=\frac13 q_1\,, \qquad q_0'=q_0\,, \qquad q_1'=-3 p^1\,.
\end{equation}
The parameters \eqref{param_prima} become exactly the ones describing the dyonic solution, \eqref{alg1}. Furthermore, the quantization relation between the charges $g_{\Lambda} p^{\Lambda} =-1$, when transformed, gives exactly:
\begin{equation}
{p^0}' {g_0}' - {g^1}' q_1'=-1\,,
\end{equation}
that matches with the condition \eqref{constr}. Furthermore, the magnetic charges are transformed in
\begin{equation}
p^0=-\frac{1}{g_0} \left(\frac14 +48 (\beta {g^1}')^2 \right)  \qquad q_1'= -3 p^1= \frac{1}{ {g^1}'}  \left(\frac34 -48 (\beta {g^1}')^2 \right)\,,
\end{equation}
that are the only electric and magnetic charges present in the dyonic configuration. Their values match with the parameterization given in \eqref{recap}.

The procedure can then be straightforwardly applied also to the non-BPS and to the nonextremal solution, and one can check that also in those cases the matching of the solutions is exact.


\begin{thebibliography}{99}



\bibitem{Sabra:1999ux}  W.~A.~Sabra,  {\it Anti-de Sitter BPS black holes in N=2 gauged supergravity},  Phys.\ Lett.\  {\bf B458 } (1999)  36-42.  [hep-th/9903143].

\bibitem{Chamseddine:2000bk}  A.~H.~Chamseddine, W.~A.~Sabra,  {\it Magnetic and dyonic black holes in D = 4 gauged supergravity},  Phys.\ Lett.\  {\bf B485 } (2000)  301-307.  [hep-th/0003213].

\bibitem{Caldarelli:1998hg}  M.~M.~Caldarelli, D.~Klemm,  {\it Supersymmetry of Anti-de Sitter black holes},  Nucl.\ Phys.\  {\bf B545 } (1999)  434-460.  [hep-th/9808097].

\bibitem{Duff:1999gh}  M.~J.~Duff, J.~T.~Liu,  {\it Anti-de Sitter black holes in gauged N = 8 supergravity},  Nucl.\ Phys.\  {\bf B554 } (1999)  237-253.  [hep-th/9901149].

\bibitem{Cucu:2003yk} S.~Cucu, H.~Lu, J.~F.~Vazquez-Poritz,{\it Interpolating from AdS$_{D-2} \times S^2$ to AdS$_D$},  Nucl.\ Phys.\  {\bf B677 } (2004)  181-222.  [hep-th/0304022].

\bibitem{Bellucci:2008cb}  S.~Bellucci, S.~Ferrara, A.~Marrani {\it et al.},  {\it d=4 Black Hole Attractors in N=2 Supergravity with Fayet-Iliopoulos Terms},  Phys.\ Rev.\  {\bf D77 } (2008)  085027.  [arXiv:0802.0141 [hep-th]].

\bibitem{Kimura:2010xe}  T.~Kimura, {\it Non-supersymmetric Extremal RN-AdS Black Holes in N=2 Gauged Supergravity}, JHEP {\bf 1009 } (2010)  061.  [arXiv:1005.4607 [hep-th]].


\bibitem{Cacciatori:2009iz}
  S.~L.~Cacciatori and D.~Klemm,
  {\it Supersymmetric AdS(4) black holes and attractors},
  JHEP {\bf 1001} (2010) 085
  [arXiv:0911.4926 [hep-th]].


\bibitem{Dall'Agata:2010gj} 
  G.~Dall'Agata and A.~Gnecchi,
  {\it Flow equations and attractors for black holes in N = 2 U(1) gauged supergravity},
  JHEP {\bf 1103}, 037 (2011)
  [arXiv:1012.3756 [hep-th]].

\bibitem{kiril_stefan}
  K.~Hristov and S.~Vandoren,
{\it Static supersymmetric black holes in AdS$_4$ with spherical symmetry},
  JHEP {\bf 1104} (2011) 047
  [arXiv:1012.4314 [hep-th]].

\bibitem{ours_nonextr}
  C.~Toldo and S.~Vandoren,
  {\it Static nonextremal AdS4 black hole solutions},
  JHEP {\bf 1209} (2012) 048
  [arXiv:1207.3014 [hep-th]].


\bibitem{klemm}
  D.~Klemm and O.~Vaughan,
  {\it Nonextremal black holes in gauged supergravity and the real formulation of special geometry},  arXiv:1207.2679 [hep-th].  

\bibitem{Larsen:1997ge}
  F.~Larsen,
  {\it A String model of black hole microstates},  Phys.\ Rev.\ D {\bf 56} (1997) 1005  [hep-th/9702153].

\bibitem{Cvetic:2010mn}
  M.~Cvetic, G.~W.~Gibbons and C.~N.~Pope,
  {\it Universal Area Product Formulae for Rotating and Charged Black Holes in Four and Higher Dimensions},  Phys.\ Rev.\ Lett.\  {\bf 106} (2011) 121301  [arXiv:1011.0008 [hep-th]].

\bibitem{Galli:2011fq}
  P.~Galli, T.~Ortin, J.~Perz and C.~S.~Shahbazi,
  {\it Non-extremal black holes of N=2, d=4 supergravity},
  JHEP {\bf 1107} (2011) 041
  [arXiv:1105.3311 [hep-th]].

\bibitem{Castro:2012av}
  A.~Castro and M.~J.~Rodriguez,
  {\it Universal properties and the first law of black hole inner mechanics},
  Phys.\ Rev.\ D {\bf 86} (2012) 024008
  [arXiv:1204.1284 [hep-th]].


\bibitem{Ceresole:2007wx}  A.~Ceresole, G.~Dall'Agata,
{\it Flow Equations for Non-BPS Extremal Black Holes},  JHEP {\bf 0703 } (2007)  110.  [hep-th/0702088].


\bibitem{ours_BPSbound}
  K.~Hristov, C.~Toldo and S.~Vandoren,
  {\it On BPS bounds in D=4 N=2 gauged supergravity},
  JHEP {\bf 1112} (2011) 014
  [arXiv:1110.2688 [hep-th]].


\bibitem{kiril}
  K.~Hristov,
  {\it On BPS Bounds in D=4 N=2 Gauged Supergravity II: General Matter couplings and Black Hole Masses},
  JHEP {\bf 1203} (2012) 095
  [arXiv:1112.4289 [hep-th]].


\bibitem{Hristov:2012nu} 
  K.~Hristov, S.~Katmadas and V.~Pozzoli,
  {\it Ungauging black holes and hidden supercharges},
  arXiv:1211.0035 [hep-th].

\bibitem{Gaillard:1981rj} 
  M.~K.~Gaillard and B.~Zumino,
  \textit{Duality Rotations for Interacting Fields},
  Nucl.\ Phys.\ B {\bf 193}, 221 (1981).

\bibitem{Andrianopoli:1996cm}
  L.~Andrianopoli, M.~Bertolini, A.~Ceresole, R.~D'Auria, S.~Ferrara, P.~Fre and T.~Magri,
  \textit{N=2 supergravity and N=2 superYang-Mills theory on general scalar manifolds: Symplectic covariance, gaugings and the momentum map},
  J.\ Geom.\ Phys.\  {\bf 23} (1997) 111
  [hep-th/9605032].

\bibitem{romans}
  L.~J.~Romans,
  {\it Supersymmetric, cold and lukewarm black holes in cosmological Einstein-Maxwell theory},
  Nucl.\ Phys.\ B {\bf 383} (1992) 395
  [hep-th/9203018].

\bibitem{batra}
  A.~Batrachenko, J.~T.~Liu, R.~McNees, W.~A.~Sabra and W.~Y.~Wen,
 {\it Black hole mass and Hamilton-Jacobi counterterms},
  JHEP {\bf 0505} (2005) 034
  [hep-th/0408205].



\bibitem{deWit:2005ub} 
  B.~de Wit, H.~Samtleben and M.~Trigiante,
  \textit{Magnetic charges in local field theory},
  JHEP {\bf 0509}, 016 (2005)
  [hep-th/0507289].


\bibitem{klemm4} 
  D.~Klemm and O.~Vaughan,
{\it Nonextremal black holes in gauged supergravity and the real formulation of special geometry II},
  arXiv:1211.1618 [hep-th].



\end{thebibliography}
\end{document}